\documentclass[aps,pra,twocolumn,showkeys,superscriptaddress]{revtex4-1}

\usepackage{graphicx}
\usepackage{dcolumn}
\usepackage{bm}
\usepackage{natbib}
\usepackage{subfigure}

\usepackage{amssymb}
\usepackage{amsmath}
\usepackage{amsthm}

\usepackage{color} 
\usepackage{cancel} 

\begin{document}


\def \Z {\mathbb{Z}}
\def \R {\mathbb{R}}
\def \C {\mathbb{C}}
\def \La {\Lambda}
\def \la {\lambda}
\def \ck {l}

\newtheorem{theorem}{Theorem}
\newtheorem{lemma}{Lemma}
\newtheorem{corollary}{Corollary}

\title{Roughness as Classicality Indicator of a Quantum State}

\author{Humberto C. F. Lemos}
\email[Corresponding author: ]{humbertolemos@ufsj.edu.br}
\affiliation{Departamento de F\'{\i}sica e Matem\'atica, CAP -
Universidade Federal de S\~ao Jo\~ao del-Rei, 36.420-000, Ouro Branco, MG, Brazil}
\author{Alexandre C. L. Almeida}
\affiliation{Departamento de F\'{\i}sica e Matem\'atica, CAP -
Universidade Federal de S\~ao Jo\~ao del-Rei, 36.420-000, Ouro Branco, MG, Brazil}
\author{Barbara Amaral}
\affiliation{Departamento de F\'{\i}sica e Matem\'atica, CAP -
Universidade Federal de S\~ao Jo\~ao del-Rei, 36.420-000, Ouro Branco, MG, Brazil}
\affiliation{International Institute of Physics, Federal University of Rio Grande 
do Norte, 59078-970, P. O. Box 1613, Natal, Brazil}
\author{Ad\'elcio C. Oliveira}
\email[Corresponding author: ]{adelcio@ufsj.edu.br}
\affiliation{Departamento de F\'{\i}sica e Matem\'atica, CAP -
Universidade Federal de S\~ao Jo\~ao del-Rei, 36.420-000, Ouro Branco, MG, Brazil}

\begin{abstract}
We define a new quantifier of classicality for a quantum state, the Roughness, which is given by the
$\mathcal{L}^2 (\R^2)$ distance between Wigner and Husimi functions.
We show that the Roughness is bounded and therefore it is a useful tool for comparison between
different quantum states for single bosonic systems. 
The state classification via the Roughness is not binary, but rather it is continuous in the interval
[0,1], being the state more classic as the Roughness approaches to zero, and more quantum when it is
closer to the unity. The Roughness is maximum for Fock states when its number of photons is arbitrarily
large, and also for squeezed states at the maximum compression limit. On the other hand, the Roughness
reaches its minimum value for thermal states at infinite temperature and, more generally, for
infinite entropy states. The Roughness of a coherent state is slightly below one half, so we
may say that it is more a classical state than a quantum one. Another important result is that the
Roughness performs well for discriminating both pure and mixed states.
Since the Roughness measures the inherent quantumness of a state, we propose another function, the
Dynamic Distance Measure (DDM),  which is suitable for measure how much quantum is a dynamics. Using
DDM, we studied the quartic oscillator, and we observed that there is a certain complementarity
between dynamics and state, i.e. when dynamics becomes more quantum, the Roughness of the state
decreases, while the Roughness grows as the dynamics becomes less quantum.
\end{abstract}


\maketitle

\section{Introduction} \label{sec:intro}

To determine if the system is classical or quantum is one of the most intriguing physics questions
of the last decades. The first challenging question was to measure the quantum state. Much effort in
this direction was made by several researchers, with many advances, both theoretical
\cite{Banaszek1996,Wallentowitz1996} and experimental \cite{Leibfried1996,Kurtsiefer1997,Banaszek1999,Bertet2002}.
The first approach to this problem was based on Ehrenfest theorem \cite%
{Ber78,Ber81,Iomin2001,Iomin2003,Berman1991,Oliveira2003,Ghose2007,Jacobs2006,Bhattacharya2000,Bhattacharya2003}
which states that, under certain conditions, the centroid of a
wave-packet state will follow a classical trajectory. Zurek and Paz
\cite{zurek1998,zurek1999} argue that the quantum system is never
isolated, and thus the dynamics of a macroscopic object is modified
by the surrounding objects that interact with it. This is the
\textit{Decoherence Approach to Classical Limit of Quantum
Mechanics}
\cite{zurek1998,zurek1999,Faria,renato2006,Oliveira09b,Oliveira2006,ZurekLosAlamos,Zur2003,adelcio2012}.
Up to our knowledge, Ballentine and collaborators
\cite{Ballentine1970,Ballentine1994,Ball1998,Ball2001,Ball2005}
where the first to address the question of which classical dynamics
would be reproduced by Quantum Mechanics, a trajectory or an
ensemble of them. Their response to this question was that in a
coarse grain approach, the quantum state may behave classically if
we consider an ensemble of trajectories.
Those results were later confirmed by others \cite%
{ZurekLosAlamos, Zur2003, Oliveira2003, Oliveira2006, Oliveira09b,
adelcio2012,Faria}. Ballentine and collaborators also argue that the
decoherence is not necessary if we take into account the
experimental limitations. This is the \textit{Coarse Grained
Approach to Classical Limit of Quantum Mechanics}. In fact, both
approaches are necessary, since there is a combination of factors that
must be considered in order to reproduce the classical regime
\cite{adelcio2012}: large actions, the interaction with the
environment and experimental observation limitations. In fact, if
Quantum Mechanics domain includes Classical Mechanics domain, then
Quantum Mechanics must reproduce all classical experiments and
observations, including individual systems like a planet or a star.
The action of the measurement apparatus on the system is closely
related to the
decoherence program \cite%
{Milburn,Romeu2012,Ghose2007,Jacobs2006,Bhattacharya2000,Bhattacharya2003},
but there is a subtle difference: if we consider a situation where the action of the environment is
negligible, the system is almost isolated, and if we perform continuous simultaneous measurements of
position and momentum, then the information about the quantum nature of the particle will be lost
and the Newtonian regime is achieved \cite{Oliveira2014,Oliveira2012b}. Those results can be
summarized in a simple way: decoherence and experimental limitations are responsible for achieving
the Liouville classical regime, while the continuous monitoring of the system leads to Newtonian
regime \cite{Oliveira2014}.

Despite the great advances on the Classical Limit problem, quantifying the degree of classicality of
a quantum state is still an open question. In the core of the Decoherence program is the assumption
that the environment is usually composed of a large number of particles, thus, due to the
thermodynamic limit, the environment (thermal bath) is essentially classical \cite{kim1996,Caldeira1983}.
Paradoxically, it has been shown that an interaction with one degree of freedom system can lead the
system to behave as it was classical, an example of a small quantum system whose classical
counterpart is chaotic and able to produces decoherence-like behavior \cite{Furuya1998}; similar
results can be found in references \cite{Bandyopadhyay2009,Casati2007,Zurek2003,Rossini2006}.
Oliveira and Magalh\~{a}es \cite{Oliveira09b} have shown that a single degree of freedom system is,
in the context of decoherence, equivalent to a $n$-degree of freedom system.
This equivalence is quantified by the effective Hilbert space size, which is ``as the Hilbert-space
size of the phase state that generates purity loss equivalently as the other particular
environmental states".
Therefore, the effective Hilbert space size is a quantifier of the effectiveness of a system as an
environment, i.e. the effectiveness of a specific model mimicking a bath is closely related to the
classicality of such state.

Given the richness of possible physical systems and the complicated structure of the quantum state
space, it is no surprise that various notions of classicality have been defined.
It seems impossible to grasp the variety of quantum states with a unique parameter, especially in
infinite-dimensional Hilbert spaces and, therefore, different classicality quantifiers should be
considered as complementary rather than competitive.

In the context of harmonic oscillator potential, many classicality quantifiers were defined in terms
of how a given state \emph{differs} from a coherent one.
These approaches follow from the postulate that coherent states are the only pure classical states
in this situation \cite{AFLP66,CN68,Hillery85}.
Some examples are Mandel Q-parameter \cite{Mandel79} and its various generalizations
\cite{Lee90, AT92, Lee94, Klyshko96}. Another approach is to use the distance of the state to the
closest classical  state defined in Ref. \cite{Hillery87}, also used in Refs.
\cite{DMMW00, WDMM01, MMS02, DR03, MMS04, Nair17}. These approaches to quantify nonclassicality
strongly depend on the chosen set of states used as reference classical states and the norms or
metrics used to define the distances. Another quantifier of nonclassicality is based on the
convolution of the $P$-function with the amount of thermal noise needed to get a non-negative
phase-space function \cite{Lee91}. Other measures are based on the entanglement potential of
non-classical states \cite{ACR05,APSM16, APPM16}. In Refs. \cite{GSV12, VS14}, the amount of
nonclassicality is quantified in terms of the minimal number of coherent states that are needed to
be superposed in order to represent the state under study. It is a member of a general class of
algebraic measures, applying to different notions of nonclassicality \cite{SV15}. A moment-based
approach was introduced to formulate measurable witnesses for the degree of nonclassicality
\cite{MSVH14}. Another approach is to determine the degree of nonclassicality based on the Fourier
transform of the Glauber-Sudarshan $P$-function, the characteristic function \cite{RSAMKHV15,RSV17}.
In reference \cite{OK12}, the authors quantify the classicality of mixed states from the perspective
of representation theory of semi-simple Lie groups and give a group theoretic characterization of
cases when it is possible to give an explicit, closed form criterion for a mixed state to be
classical. Again, the definition of classicality is heavily dependent on the criteria that coherent
states are the most classical states.

This approach can not be easily generalized to other potentials, since the coherent states of the
harmonic oscillator are not attainable for them and, therefore, cannot serve as the reference set of
classical states. The standard coherent states can be generalized for arbitrary potentials in
different nonequivalent ways \cite{NS78,GK99} and it is not clear which class of states should be
considered classical, and hence it is not clear what is the best set of reference states for the
determination of the nonclassicality of states in other potentials.

The nonclassicality of quantum states phase space is also connected with measures based on
information theory \cite{Ferraro2012,Shahandeh2017, Tan2017}. In reference \cite{Ferraro2012},
Ferraro et al. show that there are distinct notions of classicality, and, under their
considerations, that there exist quantum correlations that are not accessible by
information-theoretic arguments. 
Shahandeh et al. \cite{Shahandeh2017} show that the only known classicality criterion violated by a
non-local boson sampling protocol \cite{Aaronson2013} is the phase-space nonclassicality.
Baumgratz et al. \cite{Baumgratz2014} investigated the quantifies of resource theory of quantum
coherence. In reference \cite{Tan2017} the authors investigated non-classical light, and they show
that quantum resource \cite{Baumgratz2014} is the same of Glauber \cite{Titulaer1966}; the
non-classical light can be interpreted as a form of coherence, their procedure is based on the
negativity of $P$-distribution.

The rest of paper is organized as follows: in section \ref{sec:defs} we define a new measure, the
Roughness $R$, and prove that it is bounded between $[0,1]$. Given two states, we say that the one with
a larger value of $R$ is more non-classical than the other. In section \ref{sec:exm} we address some
important quantum states, and evaluate the Roughness for each one of them. We stress that we could find,
both for lower and upper bounds, examples of states that, in limit case, achieve those values. In
section \ref{sec:neg} we compare the Roughness with another classicality measure, the Negativity $N$.
First, $N$ is not a bounded function, so it can be more difficult to compare any two given states.
Also, we show that there are some states with $N=0$ (said to be totally classical), but with $R>0$,
i.e. the Roughness can find some quantumness in such cases.
Particularly, we study a convex mixing between a thermal and a Fock state, and
supported by entropy, we show that the Roughness is more reliable, especially for small temperatures.
At last, in section \ref{sec:DDM} we define another classicality measure, the Dynamic Distance
Measure $D$.
While $R$ evaluates the inherent quantumness of a state, DDM quantifies how much a quantum dynamics
is far from a classical one.
We numerically evaluate both $R$ and $D$ for the quartic model, and we find a complementary behavior
between them for such model.

\section{Roughness: definition and bounds} \label{sec:defs}

The Wigner quasipropability distribution, better known as Wigner function, was introduced in 1932
by Eugene Wigner \cite{Wigner1932}.
It is a real-valued function for any arbitrary quantum state $\Psi$, and it is given by
\begin{equation} \label{eq:W}
W_{\Psi}(q,p) = \frac{1}{2\pi} \int_{\R} dx\, e^{ipx}
 \left\langle q-\frac{x}{2} \Big| \Psi \right\rangle
\left\langle \Psi \Big| q+\frac{x}{2} \right\rangle .
\end{equation}
As a distribution, it is normalized, i.e.
$$
\int_{\R^2} dq\, dp\, W_{\Psi}(q,p)=1\, ,
$$
it is also common to say that it has unitary volume. The Wigner function is a real bounded function,
with $|W_{\Psi}(q,p)| \leq \pi^{-1}$ for any $(q,p) \in \R^2$. Moreover, it is square integrable
\begin{equation} \label{eq:W2}
\|W_{\Psi}\|^2 = (W_{\Psi},W_{\Psi}) = \int_{\R^2} \!\!\!\!
dq\, dp\, [W_{\Psi}(q,p)]^2 \leq \frac{1}{2\pi}\, ,
\end{equation}
and the equality above holds when $\Psi$ is a pure state \cite{Leonhardt2010}.
The inner product above is the canonical one in the $\mathcal{L}^2(\R^2)$ space. Among its
properties, we emphasize the fact that $W_{\Psi}(q,p)$ can assume negative values, so it cannot be a
regular probability distribution, and for this reason it is seen as a signature of quantumness of
the state. Actually, there is a measure for non-classicality based on this property, namely the
Negativity \cite{Kenfack2004}
\begin{eqnarray}
N (\Psi) &=& \int_{\R^2}  \Big[ |W_{\Psi}(q,p)| - W_{\Psi}(q,p) \Big] dq\, dp =
\nonumber \\
 \label{eq:N}
&=& \int_{\R^2} |W_{\Psi}(q,p)| dq\, dp - 1 ,
\end{eqnarray}
which evaluates the volume of the negative part for Wigner function.
The classicality quantifier $N$ above should not be confused with the negativity measure for
entanglement \cite{Vidal2002}.

Inspired by this same property for Wigner function, we propose here a new measure of how quantum is
a state. First we have to refer to another distribution, the Husimi function or $Q$-distribution
\cite{Leonhardt2010}, which can be evaluated from Wigner function as
\begin{equation} \label{eq:Hu}
Q_{\Psi}(q,p)=\frac{1}{\pi }\int_{\R ^{2}} \!\!\!\!
dq^{\prime }\,dp^{\prime }\,W_{\Psi}(q^{\prime
},p^{\prime}) e^{-[(q-q^{\prime})^2+(p-p^{\prime} )^2]}\, .
\end{equation}%
In other words, we convolute the $W$ function for any given state with the Gaussian distribution for
a vacuum state. It smooths the oscillations of the Wigner function around a point $(q,p)$ in the
phase space, as we average $W(q',p')$ values in a circle around this point.
Consequently, one can show that $Q_{\Psi}(q,p)$ is always non-negative, for any $\Psi$, so it
is always acceptable as a classical distribution -- other quasi-distributions, like the $P$-function
\cite{Leonhardt2010}, do not have such property.
For these reasons, we have chosen Wigner and Husimi functions to propose a new way of measuring
non-classicality.
The $R$ measure, namely the Roughness, was inspired by standard measures of roughness
\cite{BlackKohser201108} and is defined as proportional to the $\mathcal{L}^2 (\R^2)$ distance
between both functions. Our idea is that Wigner function, as discussed before, carries some very
important information of the quantumness for a given state and since $Q$ smooths the oscillations of
$W$, and also it can be seen as a regular probability distribution, we can quantify how much
non-classical a state is by evaluating how far those functions are one from another.
The Roughness is given by
\begin{eqnarray} \label{eq:R}
R(\Psi) &=& \sqrt{2\pi} \, \big\| W_{\Psi} - Q_{\Psi} \big\| =
\\
&=& \sqrt{ 2\pi \int_{\R^2} \!\!\!\! dq \, dp\, \big| W_{\Psi} (q,p) - Q_{\Psi} (q,p) \big|
^{2} }  .
\nonumber
\end{eqnarray}
We classify a state as more non-classical when its Wigner function is more distant form its Husimi
function. 
In other words, given two any states, the one with larger $R$ will be more quantum. From now on, for
the sake of simplicity, we drop the $\Psi$ index on $W$ and $Q$ functions notation, unless it is
necessary to make it clear.

As a first property, we show that $R$ is bounded
\begin{equation} \label{eq:Rbounds}
0 \leq R \leq 1 ,
\end{equation}
for any state. The lower bound is obvious from definition (\ref{eq:R}). We now prove the upper
bound. First we define the symmetric Fourier transform for Wigner function
\begin{equation} \label{eq:What}
\widehat{W}(u,v)=\frac{1}{2\pi }\int_{\R^{2}}dq\,dp\,e^{-i(uq+vp)}W(q,p)\, ,
\end{equation}
and in an analogous way, the Fourier transform $\widehat{Q}(u,v)$. If we name the Gaussian function
$g(q,p) = e^{-(q^2+p^2)}$, we can see from (\ref{eq:Hu}) that the Husimi function is merely the
convolution
$$
Q(q,p) = 2 ( W \ast g) (q,p)\, .
$$
Thus we have
\begin{equation} \label{eq:Qhat}
\widehat{Q}(u,v)=\exp \left( -\frac{u^{2}+v^{2}}{4}\right) \widehat{W}(u,v)\,.
\end{equation}%
Using Plancherel theorem \cite{ReedSimon1975}, we obtain
\begin{eqnarray*}
R^{2} &=&2\pi \Vert W-Q\Vert ^{2} = 2\pi \Vert \widehat{W}-\widehat{Q}\Vert ^{2} \\
&=&2\pi \int_{\R^{2}}du\,dv\,\left( 1-e^{-\frac{1}{4}(u^{2}+v^{2})}\right)
^{2}\left\vert \widehat{W}(u,v)\right\vert ^{2} \\
&\leq &2\pi \int_{\R^{2}}du\,dv\,\left\vert \widehat{W}(u,v)\right\vert
^{2}=2\pi \Vert \widehat{W}\Vert ^{2}=2\pi \Vert W\Vert ^{2}\leq 1,
\end{eqnarray*}%
where we have used (\ref{eq:W2}) in the last step. As a consequence, given any state $\Psi$, its
Roughness $R(\Psi)$ will be always bounded. The Roughness will be closer to one as more
non-classical a state is; on the other hand, if a state is more classical, its Roughness will be
closer to zero. We will show on section \ref{sec:exm} that there are, for both bounds, states which
can be arbitrarily close to these values.

\section{Examples} \label{sec:exm}

We now evaluate the Roughness of some common and important quantum states \cite{Leonhardt2010} that
will give us some insights about $R$. Also, these states appear in many applications in Quantum
Optics \cite{Dodonov2002} and other areas.

\subsection{Coherent state}

A coherent state is a specific quantum state of the quantum harmonic oscillator, often described as
a state whose dynamics most closely resembles the oscillatory behavior of a classical harmonic
oscillator. Its Wigner and Husimi functions respectively are
\begin{eqnarray} \label{eq:W0}
W_0(q,p) &=& \frac{1}{\pi} e^{-(q^2+p^2)},
\\
Q_0(q,p) &=& \frac{1}{2 \pi} e^{-\frac{q^2+p^2}{2}}.
\nonumber
\end{eqnarray}
It is straightforward to evaluate its roughness as
\begin{equation} \label{eq:R0}
R_0 = \frac{1}{\sqrt{6}} \approx 0.408 \, .
\end{equation}
Based on our numerical investigations, we conjecture that this is the smallest value for the
Roughness of a pure state, and we use this value as a reference to compare to other states.
Also, we can see that the coherent state is roughly in the middle of the Roughness scale. Since its
Roughness is just below one half, we can say that the coherent state is closer to classical than
to quantum, but as it is a pure state, its entropy is zero and therefore its Roughness is greater
than that of other non-pure states.

We must emphasize that  (\ref{eq:W0}) is the Wigner function for the coherent state centered at the
origin, but if we center it in another any point $(q_0,p_0)$, we get
$$
W_{(q_0,p_0)} (q,p) = W_0(q-q_0,p-p_0) = \frac{1}{\pi} e^{-[(q-q_0)^2+(p-p_0)^2]}.
$$
We can easily show, by using Fourier transform properties, that such state has the same Roughness
$R_0$ (\ref{eq:R0}).

\subsection{Harmonic Oscillator eigenstates: Fock states} \label{ssec:Fock}

A Fock state is an eigenstate of the number operator with eigenvalue $n$, and it is a pure state
$\rho_n = |n\rangle \langle n|$.
The Fock state has been measured in many physical contexts: in a superconducting quantum circuit
\cite{Hofheinz2008}, in superconducting quantum cavity \cite{Bertet2002}, and in the context of trapped
ions \cite{Jose2000}.
Its Wigner function has been investigated experimentally, and it is negative at some points, as
shown in \cite{Bertet2002}. Indeed, it is given by
\begin{equation} \label{eq:WFock}
W_{n}(q,p)=\frac{(-1)^{n}}{\pi }e^{-(q^{2}+p^{2})}L_{n}\left(
2(q^{2}+p^{2}) \right) \ ,
\end{equation}
where $L_n(x)$ is the known Laguerre polynomial of order $n$ \cite{Arfken2011}. The polynomial $L_n$
has $n$ strictly positive real zeros \cite{Szego1939}. Moreover, they are all in a finite open
interval $(0,\nu)$, where this upper bound $\nu$ is well known \cite{Gatteschi2002}. It means that
$L_n(x)$ has an oscillatory part on this interval, so it changes its signal $n$ times, which suggests
that, as larger as $n$ gets, the negative part of $W_n$ becomes more significant, so its Roughness
increases: our results confirm such insight.
The Husimi function for a Fock state is \cite{Leonhardt2010}
\begin{equation}
Q_{n}(q,p)=\frac{1}{2\pi \,n!}\left( \frac{q^{2}+p^{2}}{2}\right) ^{n}e^{-%
\frac{1}{2}(q^{2}+p^{2})}\ .  \label{eq:HFock}
\end{equation}
Taking $n=0$ in both (\ref{eq:WFock}) and (\ref{eq:HFock}), we obtian the coherent state
(\ref{eq:W0}).

Our calculations for the Roughness are tedious, but straightforward. We give more details on
\ref{app:Fock}. By definition, the Roughness is
\begin{eqnarray}
R_{n}^{2} &=&2\pi \int_{\mathbb{R}^{2}}dq\,dp\,\big[ W_{n}(q,p)-Q_{n}(q,p)\big] ^{2} =
\nonumber
\\
\label{eq:R2termos}
&=& R_{W_{n}^{2}}^{2}+R_{Q_{n}^{2}}^{2}-R_{W_{n}Q_{n}}^{2} \, ,
\end{eqnarray}
where we have defined 
\begin{subequations}
\begin{eqnarray}
 \label{eq:R2W2def}
R_{W_{n}^{2}}^{2} &:=& 2\pi \int_{\mathbb{R}^{2}}dq\,dp\, \big[ W_{n}(q,p)\big] ^{2},
\\
 \label{eq:R2Q2def}
R_{Q_{n}^{2}}^{2} &:=& 2\pi \int_{\mathbb{R}^{2}}dq\,dp\, \big[ Q_{n}(q,p)\big] ^{2},
\\
 \label{eq:R2WQdef}
R_{W_{n}Q_{n}}^{2} &:=& 4\pi \int_{\mathbb{R}^{2}}dq\,dp\, \big[ W_{n}(q,p) - Q_{n}(q,p) \big]. 
\end{eqnarray} 
\end{subequations}
We find -- details on \ref{app:Fock} -- that
\begin{subequations}
\begin{eqnarray} \label{eq:R2W2}
R_{W_{n}^{2}}^{2} &=& 1 , \qquad \forall n =0,1,2,\ldots ,
\\
R_{Q_{n}^{2}}^{2} &=& \frac{1}{2^{2n+1}} {2n \choose n}
= \frac{1}{2} \frac{(2n)!}{2^{2n}(n!)^2} > 0 \, \label{eq:R2H2},
\\
R_{W_{n}Q_{n}}^{2} &=& \frac{4}{3}\left( -\frac{1}{3}\right)^{n} F\left(
-n,n+1;1;\frac{4}{3}\right) =
\nonumber\\
&=& \frac{4}{3}\left( -\frac{1}{3}\right)^{n} \sum_{j=0}^{n} \frac{(n+j)!}{(j!)^2(n-j)!}
\left( - \frac{4}{3} \right)^{j} \, ,  \label{eq:R2WH_F}
\end{eqnarray}
\end{subequations}
where $F$ is the hypergeometric function, which becomes a finite sum if either its first or second
argument is a negative integer, as it happens on (\ref{eq:R2WH_F}). The results
(\ref{eq:R2W2})-(\ref{eq:R2WH_F}) were obtained using some known integrals for Laguerre functions
\cite{Gradshteyn2007}.
It is not straightforward to see on the equations above, but one can check that for $n=0$ we recover
$R_0^2=1/6$, as expected.
Also, we emphasize that $R_{W_{n}^{2}}^{2} = 1$ for any $n$-Fock state, and it is a typical
characteristic for pure states, as we have already said just after Eq. (\ref{eq:W2}).

\begin{figure}[h]
\center \subfigure[ref1][Roughness of a Fock state as function of
$n$.]{\includegraphics[width=\linewidth]{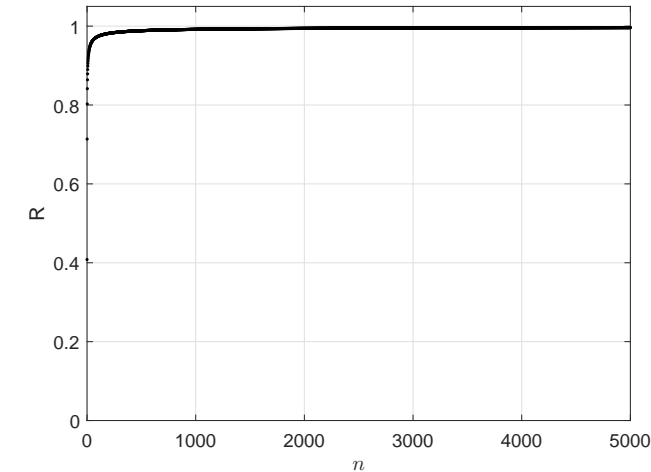}}
\qquad \subfigure[ref2][Magnification of
(a).]{\includegraphics[width=\linewidth]{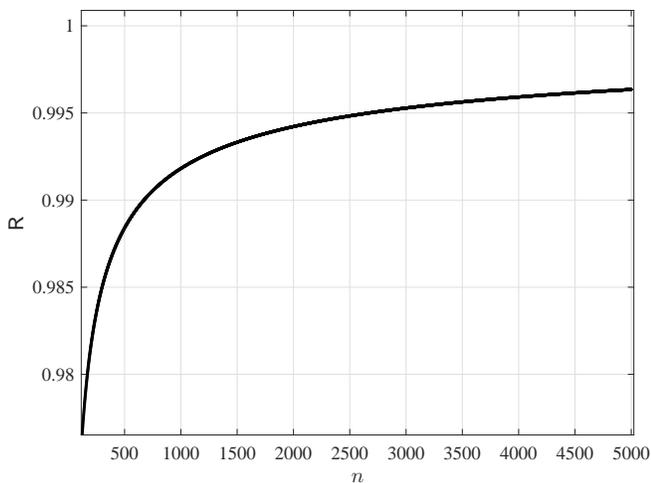}}
\caption{Roughness for the Fock state $|n\rangle$ as function of
$n$.} \label{Fock}
\end{figure}

We also prove in \ref{app:Fock} that
\begin{equation} \label{eq:ineq}
0 < R_{Q_{n}^{2}}^{2} < R_{W_{n}Q_{n}}^{2}, \quad \forall n .
\end{equation}
The inequality above is important: replacing it in (\ref{eq:R2termos}), we can check that our upper
bound (\ref{eq:Rbounds}) is respected, as it should be.
It is hard to see property (\ref{eq:ineq}) from Eq. (\ref{eq:R2WH_F}), but on \ref{app:Fock} we
rewrite this term as
\begin{equation} \label{eq:R2WH_2}
R^2_{W_nQ_n} = \frac{4}{3} \left( \frac{1}{9} \right)^n \sum_{j=0}^n {n \choose j}^2 4^j >0.
\end{equation}
Moreover and more important, we also prove that
\begin{equation} \label{eq:limits}
\lim_{n \to \infty}R_{Q_{n}^{2}}^{2} = 0 =
\lim_{n \to \infty}R_{W_{n}Q_{n}}^{2} ,
\end{equation}
and so we have for the Roughness for the Fock state that
\begin{equation} \label{eq:Rn}
\lim_{n \to \infty}R_n = 1.
\end{equation}
This result is quite remarkable: the Roughness for the Fock state increases as $n$ becomes larger,
and it reaches the upper bound on the limit $n \to \infty$. In other words, as $n$ increases, the
Fock state $\rho_n$ becomes more non-classical, and it gets arbitrarily closer to maximum value for
the Roughness for a sufficiently large $n$. This result contradicts those who argue that the Fock state
becomes more classical as $n$ increases -- see, for example, \cite{Home2013} and references therein.
In figure \ref{Fock}, we show the Roughness of a Fock state dependence on $n$. Although the Fock
state approaches the Roughness upper bound, the convergence to unity is very slow.

As a last comment for this subsection, Fock states can be used as basis for more general states
\begin{equation} \label{eq:FockGen}
\rho  = \sum_{n,m=0}^{\infty} A_{n,m} \left| n \right\rangle \left\langle m \right| ,
\end{equation}
where, of course, $A_{n,m}$ are complex constants such that $\mathrm{Tr}\, \rho =1$, and
$\rho^{\dagger} = \rho$.
In  \ref{app:FockGen} we give detailed analytical results that are useful for evaluating the Roughness
for these states (\ref{eq:FockGen}).

\subsection{Squeezed states}

The squeezed states were presented in 1927, by Kennard \cite{Kennard1927}, as the first example of
non-classical states. The Wigner function for a squeezed state is
\begin{equation}\label{eq:squeezed}
W_{\zeta}(q,p)=\frac{1}{\pi }\exp \left[- (e^{2\zeta}q^{2}+e^{-2\zeta}p^{2}) \right] ,
\end{equation}
where $\zeta \in \R$. 
Actually, we are only considering states which are squeezed along the principal axes, a more general
Gaussian state would take rotations and translations into account.
Once again, if $\zeta=0$, we have the coherent state (\ref{eq:W0}).
If $\zeta>0$, we have a narrower Gaussian in $q$ and a wider one on $p$, the opposite happens for
$\zeta<0$.
It is a consequence of Heisenberg's uncertainty principle: a narrower Gaussian in $q$ means that we
have a larger probability that our state is localized on a small neighborhood of $q=0$, but as a
consequence, the wider Gaussian in $p$ tells us that we have a significant probability to have any
velocity. Intuitively, we think that as large as $|\zeta|$ gets, the squeezed state becomes more
non-classical, and our results corroborate this statement.

The Fourier transform for its Wigner function is
$$
\widehat{W}_{\zeta}(u,v)=\frac{1}{2\pi }\exp \left[ -\frac{1}{4} (e^{-2\zeta}u^{2}+ e^{2\zeta}v^{2}) \right] ,
$$
and so, from definition, we can straightforward obtain
\begin{equation} \label{eq:RSqueezed}
R(\zeta)=\left[ 1+\frac{e^{\zeta}}{e^{2\zeta}+1}-\frac{4e^{\zeta}}{\sqrt{(e^{2\zeta}+2)(2e^{2\zeta}+1)}}
\right]^{\frac{1}{2}} \, .
\end{equation}
We can easily check that $R(0)= 1/\sqrt{6}$. Also, we can prove that
\begin{equation} \label{eq:Rsqlim}
\lim_{\zeta \to \pm \infty} R(\zeta) = 1,
\end{equation}
so it is another example of a state that reaches the maximal quantumness in our Roughness measure.

\begin{figure}[h]
\includegraphics[width=\linewidth]{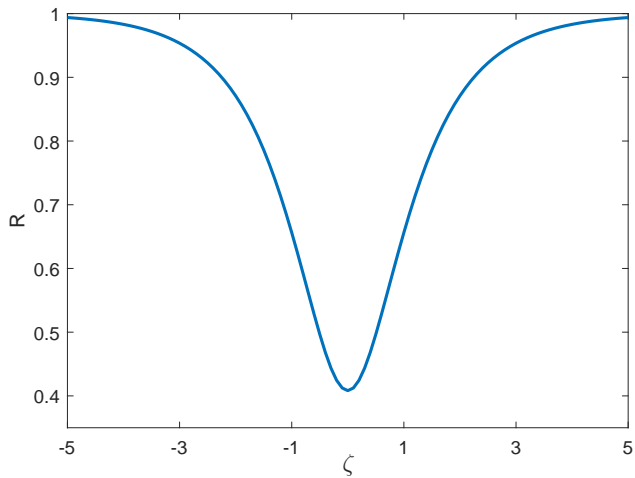}
\caption{Roughness of the squeezed state as function of $\zeta$.}
\label{Sqz1}
\end{figure}

In figure \ref{Sqz1} we show how the Roughness for squeezed states depends on $\zeta$. Minimum
Roughness occurs at $\zeta=0$, and in this case, the squeezed state is just a coherent one.

\subsection{Field cat states} \label{ssec:cat}

Field cat states are usually known as the most common example of a non-classical state
\cite{Dodonov2002}. They are given by the superposition of two coherent states,
$  \left| \alpha e^{\pm i\phi } \right\rangle,$
\begin{equation*} 
\left| C\right\rangle (\phi ,\alpha )=\mathcal{N}%
\left( \left| \alpha e^{i\phi }\right\rangle +(-1)^{n}\left| \alpha
e^{-i\phi }\right\rangle \right),
\end{equation*}%
where $\mathcal{N}$ is the normalization constant, $\phi$ is the phase, and $\alpha e^{\pm i\phi }$
determines the center of the coherent state.
For $\alpha = -iq_0$ and $\phi=\pi/2$, we have
\begin{equation}  \label{eq:cat}
\left| C\right\rangle (\phi ,\alpha )=  \frac{\left| q_{0}\right\rangle \pm \left| -q_{0}
\right\rangle}{\sqrt{2}} \, ,
\end{equation}%
where the plus (minus) sign refer to even (odd) cat state. We highlight that odd cat state is not
defined for $q_{0}=0$; on the other hand, if we take $q_{0}=0$ for even cat state, we recover the
coherent state related to Wigner function (\ref{eq:W0}).
Indeed, the Wigner functions for (\ref{eq:cat}) are
\begin{widetext}
\begin{equation*} \label{eq:Wcat}
W_{\pm} (q,p) = \frac{e^{-[(q-q_0)^2+p^2]} + e^{-[(q+q_0)^2+p^2]} \pm
2 e^{-(q^2+p^2)} \cos (2q_0p)}{2\pi (1\pm e^{-q_0^2})} \, .
\end{equation*}
\end{widetext}

Roughness calculations can be easily done by using Fourier transform with its properties and eq.
(\ref{eq:Qhat}). We obtain
\begin{equation} \label{eq:Rpar}
R^2_{+} (q_0) = \frac{7}{12} + \frac{e^{-q_0^2}}{(1+e^{-q_0^2})^2} -
\frac{2}{3} e^{-\frac{2}{3}q_0^2} \left( \frac{1+e^{-\frac{1}{3}q_0^2}}{1+e^{-q_0^2}}  \right)^2 ,
\end{equation}\begin{equation} \label{eq:Rimpar}
R^2_{-} (q_0) = \frac{7}{12}   -
\frac{2}{3} e^{-\frac{2}{3}q_0^2} \left( \frac{1-e^{-\frac{1}{3}q_0^2}}{1-e^{-q_0^2}}  \right)^2 .
\end{equation}
First, we can easily see that
$
R_{+} (0) = \sqrt{ 1/6},
$
as expected. Moreover, we get
\begin{equation} \label{eq:catlimit}
\lim_{q_0 \to \pm \infty} R_{+} (q_0) = \lim_{q_0 \to \pm \infty} R_{-} (q_0) =
\sqrt{\frac{7}{12}} \approx 0.764\, .
\end{equation}

\begin{figure}[h]
{\includegraphics[width=\linewidth]{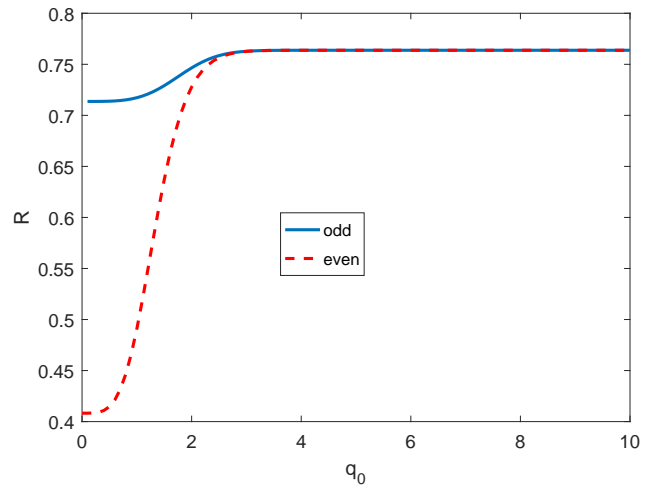}}
\caption{Roughness for the odd (blue line) and even (red dashed
line) cat states as function of $q_0$, as given by eq. (\ref{eq:cat}).}
\label{fig:cat}
\end{figure}

The result above shows us, alongside with figure \ref{fig:cat}, that the Roughness for odd cat state
is always larger than the even one, and both become more non-classical as $q_0$ increases, but not
even on limit they reach the maximum value for the Roughness.
As we increase $q_0$, from $q_0\approx 2$ (for the even cat) or $q_0>0$ (for the odd cat), the cat
state Roughness is significantly greater than the coherent state, thus corroborating the statement
that cat state is non-classical.
However, we emphasize that no matters how large we take $q_{0}$ for a cat state, we always get a
more quantum Fock state (for large $n$) or squeezed state (for large $|\zeta|$).
In other words, our results show the cat states as non-classical ones, but there are states which
are ``more quantum".

\subsection{Thermal state}

A thermal state for the Harmonic Oscillator with frequency $\omega$ is a mixed state given by
\begin{equation} \label{eq:Therm}
\rho _{T}=\left(1-e^{-\frac{\hbar \omega }{k_{B}T}}\right)\sum_{n=0}^{\infty }e^{-\frac{%
\hbar \omega }{k_{B}T}n}\left\vert n\right\rangle \left\langle
n\right\vert
\end{equation}
where $\hbar $ and $k_{B}T$ have their usual meaning, and $\left| n\right\rangle $ is the harmonic
oscillator eigenstate (the Fock state). Its Wigner and Husimi functions are
\begin{eqnarray} \label{eq:WTherm}
W_{\bar{n}} (q,p) &=& \frac{1}{\pi(2\bar{n}+1)} \exp\left(
-\frac{q^2+p^2}{2\bar{n}+1} \right) ,
\\
Q_{\bar{n}} (q,p) &=& \frac{1}{2\pi(\bar{n}+1)} \exp\left(
-\frac{q^2+p^2}{2(\bar{n}+1)} \right) , \nonumber
\end{eqnarray}
where $\bar{n}=\left(e^{\frac{\hbar \omega}{k_BT}}-1\right)^{-1}$ is the thermal average number of
photons in a mode. For $\bar{n}=0$ we have the coherent state (\ref{eq:W0}) again. Also one can see
that, for $\bar{n}\to \infty$, $W_{\bar{n}}$ and $Q_{\bar{n}}$ have the same limit, so we can expect
that the Roughness goes to zero as $\bar{n}$ increases. It is straightforward to evaluate the
Roughness for the thermal state as
\begin{equation} \label{eq:RTherm}
R_{T}(\bar{n}) = \left[ \frac{1}{2}
\frac{1}{(\bar{n}+1)(2\bar{n}+1)(4\bar{n}+3)}
\right]^{\frac{1}{2}}\, .
\end{equation}
As expected, if $\bar{n}=0$ we recover $R_0$ (\ref{eq:R0}). Also, we can easily see that
$R_{T}(\bar{n}) \to 0$, as $\bar{n} \to \infty$, and it is a consequence of the fact that the
quantum partition function becomes closer to the classical one as $\bar{n}\to \infty$.
Moreover, we can see that it goes to zero as $\bar{n}^{-3}$.

\subsubsection{Diagonal State}

Another non pure state is the Diagonal state of order $(m+1)$, which is defined as
\begin{equation}
\rho _{D}(m)=\frac{1}{m+1}\sum_{n=0}^{m}\left\vert n\right\rangle
\left\langle n\right\vert .
\end{equation}%
The diagonal state represents a mixed state with uniform distribution. It is easy to see that in the
limit case $m\to \infty$, we will have $R\to 0$, since $R$ is bounded.

The mean photon number for the Diagonal State is $ \bar{n}=\mathrm{Tr} \, {[\hat{N}\rho _{D}(m)]}
=m/2$, where $\hat{N}$ is the number operator.
Again, as in the thermal state case, the Roughness goes to zero when the mean photon number goes to
infinity, but in this case, the convergence is slower. The Roughness of the Diagonal state was
determined numerically using the results of section \ref{ssec:Fock} and \ref{app:FockGen}.

\begin{figure}[htbp]  \label{termico}
{\includegraphics[width=\linewidth]{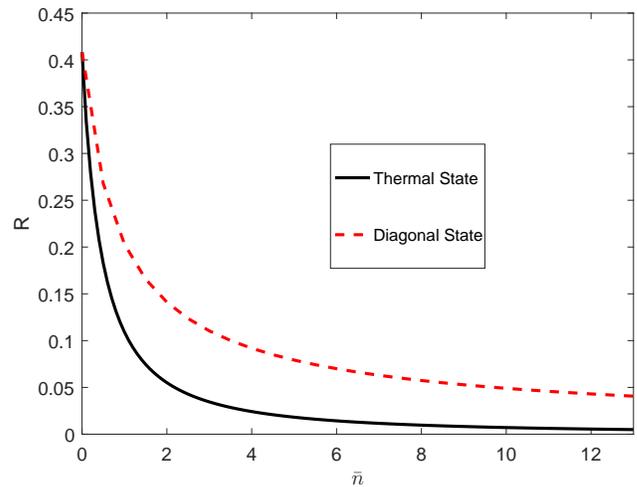}} 
\caption{Roughness for a Thermal state (black line) and Roughness
for the Diagonal state (red dashed line) as function of mean photon number
$\bar{n}$.}
\end{figure}

Now we compare some features for thermal and Diagonal states. In figure \ref{termico}, we show the
Roughness for both states as function of $\bar{n}$. We can see that even for small values of
$\bar{n}$, $R$ is already close to zero, which means that it would be very difficult to observe
quantum features in these states.

\begin{figure}[htbp]
{\includegraphics[width=\linewidth]{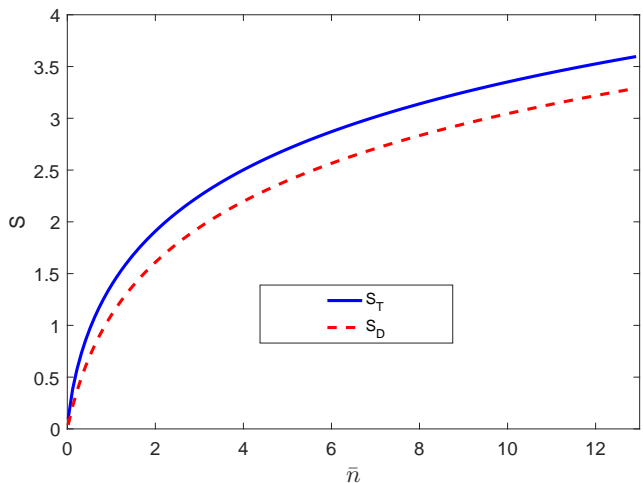}} 
\caption{Entropy for a Thermal state (black line) and Entropy for
the Diagonal state (red dashed line) as function of mean photon number
$\bar{n}$.} \label{entropy}
\end{figure}

The difference between thermal and Diagonal states is one of those situations where the observable
choice determines the system classicality measure. Since the states are not pure, the entropy can be
used to quantify their purity. If we consider the same $\bar{n}$ for both states, we have different
values for the Roughness, but the linear entropy (defined as $\delta =1-\mathrm{Tr} \, \rho^2$) is
identical for both states, and it is given by
\begin{equation*}
\delta = 1-\mathrm{Tr}\, (\rho _{D}^{2})
= 1-\mathrm{Tr}\, (\rho _{T}^{2})
=\frac{2\bar{n}}{2\bar{n}+1}.
\end{equation*}
Although they have the same linear entropy, they do not have the same entropy. The entropy for
thermal state is
\begin{equation*}
 S_{T}=k_{B}(1+\bar{n})\ln \left[ 1+\overline{n}\right]
-\bar{n}k_{B}\ln \left( \overline{n}\right),
\end{equation*}
while the entropy for Diagonal state is
\begin{equation*}
S_{D}=k_{B}\ln (2\bar{n}+1).
\end{equation*}
As we can see in figure \ref{entropy}, for the same mean photon number $\bar{n}$, thermal state has
a bigger entropy, which explains why its Roughness is smaller. The Roughness is more sensitive to
the difference between thermal and diagonal states than Linear entropy.
This result is an example that the classicality of a system is sensible to which criteria is used to
quantify it, and not only on the observable choice \cite{Oliveira2003}, since both entropy and
linear entropy can be estimated by the same set of measurements.

\section{Roughness $\times$ Negativity: a comparative study}
\label{sec:neg}

Negativity is widely used as a measure of classicality for a quantum state \cite{Kenfack2004}, and
it is defined by Eq. (\ref{eq:N}). In this section, we compare both quantifiers.
First, we remark directly from its definition that Negativity is zero for any state whose Wigner
function is positive, i.e. it does not distinguish a thermal state from a coherent state and/or
squeezed state, while Roughness can do the trick, as we have seen on section \ref{sec:exm}.
Another question is about bounds for $N$, since up to our knowledge it is not a bounded
function -- actually, it is shown in \cite{Kenfack2004} that the Negativity grows proportionally to
$n^{1/2}$ for Fock states, at least for $n$ up to 250 --, and there are some results on the integral
of the Wigner function over a sub-region of the phase space of a one degree of freedom quantum system
which can be less than zero or greater than one on this sub-region \cite{Bracken}. The lack of known
bounds for Negativity can be a problem if one needs to compare different states.

Now, in order to clarify other advantages of Roughness over Negativity, we study a mixture state
$\rho _{z}$ given by the convex combination
\begin{equation} \label{termicoz}
\rho _{z}(\beta ,M)=(1-z)\rho _{\beta }+z\left\vert M\right\rangle
\left\langle M\right\vert ,
\end{equation}%
\newline
where $\rho _{\beta }$ is the thermal state (\ref{eq:Therm}) at temperature
$T=1/\beta k_{B}$, and $\left\vert M\right\rangle $ is a $M$ Fock state.
We recall here that, although the mixed state (\ref{termicoz}) is a linear combination between a
thermal and a Fock state, and so are their respective Wigner and Husimi functions, its Roughness,
directly from the definition (\ref{eq:R}), is nonlinear on $z$.
Moreover, thermal states have positive Wigner functions (\ref{eq:WTherm}), so $N$ is always zero for
them, while $R$ is not, as we can see in Eq. (\ref{eq:RTherm}).
The idea of studying such mixture state is that a Fock state $|M\rangle$ is always a pure state, but
it can be as quantum as we want, as we have shown in eq. (\ref{eq:Rn}).
Alternatively, $\rho _{\beta}$ is a pure state only in the limit $T \to 0^{+}$, namely the
coherent state. As the temperature increases,  $\rho _{\beta}$ becomes more non-pure, while its
Roughness goes to zero (\ref{eq:RTherm}). So, for sufficiently large values of $\beta$, we are
practically mixing two pure states, but Roughness for $\rho _{\beta}$ is given by $R_0$
(\ref{eq:R0}), while we can take a large $M$ to get a Fock state whose Roughness is as close to the
unity as we want to.
On the other hand, for small values of $\beta$, we have the same Fock state $|M\rangle$, but
$\rho _{\beta}$ is more non-pure and more classical as $T$ gets larger.

\begin{figure}[htb]
\includegraphics[width=\linewidth]{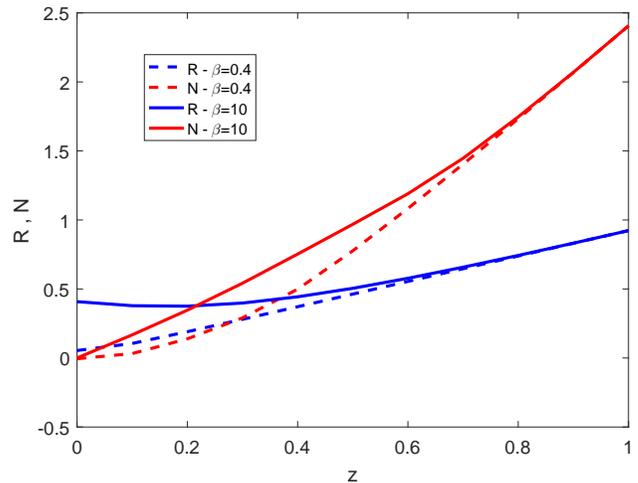}
\caption{Roughness (blue) and Negativity (red) as function of $z$
for the state $\rho_{z}$ for $M=10$, and temperatures $\beta=0.4$ (dashed lines) and
$\beta=10$ (full lines).}
\label{fig:RxN}
\end{figure}

\begin{figure}[htb]
\includegraphics[width=\linewidth]{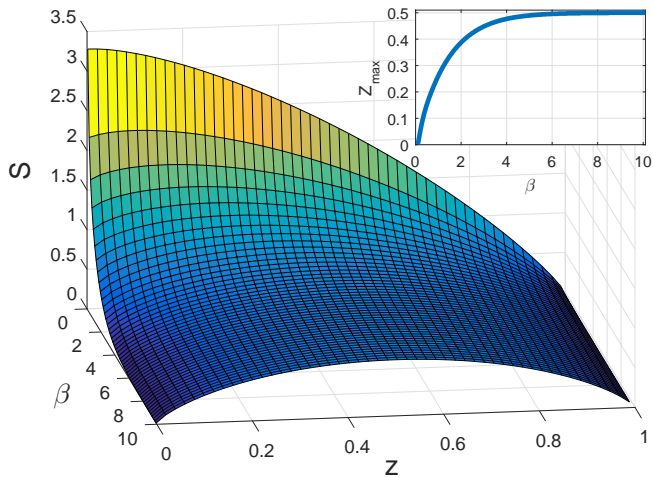}
\caption{Entropy as function of temperature $\beta$ and convex combination parameter $z$.
On the inset we plot $z_{max}$ -- the value of $z$ where $S$ is maximum -- dependence on
$\beta$.}
\label{fig:SXzXbeta}
\end{figure}

In such spirit, we plot in figure \ref{fig:RxN} Roughness and Negativity for $M=10$ (taken as large
$M$), both for small $\beta = 0.4$ (doted line) and large $\beta = 10$ (full line).
In any case, we expect that $R(z=1) > R(z=0)$, since $R_0$ is at the same time a lower bound for
Roughness for a Fock state and an upper bound for a thermal state. Nonetheless, for large $\beta$ we
have an almost pure state when $z=0$ and a genuine pure state when $z=1$, and pure states are
typically quantum. So, for small $0<z<1$, we might expect that this mixture becomes less quantum,
and we can clearly see it in figure \ref{fig:RxN}, as $R(\beta=10)$ in a non-monotonic function
of $z$. For small $\beta$ the initial $z=0$ thermal state is already non-pure, and for this reason
$R$ is monotonic in $z$ in this case. Facing this behavior, the Negativity is always monotonic in
$z$, as $N=0$ for thermal states and $N>0$ for any Fock state such that $M \geq 1$.
Our results for Roughness for the mixed state (\ref{termicoz}) are, once again, supported by
entropy. Indeed, on figure \ref{fig:SXzXbeta} we plot the entropy $S$ as function on $z$ and
$\beta$. We can see that, for small fixed temperatures (large $\beta$), that the entropy is close to
zero -- as we said, for $z=0$ the thermal state is almost pure in such case.
So $S$ clearly increases as we take small values for $z>0$, and after $S$ reaches its maximum, it
decreases to $S=0$ at $z=1$, since now we have a pure Fock state.
However, for large temperatures, we already start from a very entropic state, so the entropy is
large for $z=0$. In the inset of figure \ref{fig:SXzXbeta} we show $z_{max}$, the value of $z$ where
entropy $S$ is maximum, as function of $\beta$.

\begin{figure}[htb]
\includegraphics[width=\linewidth]{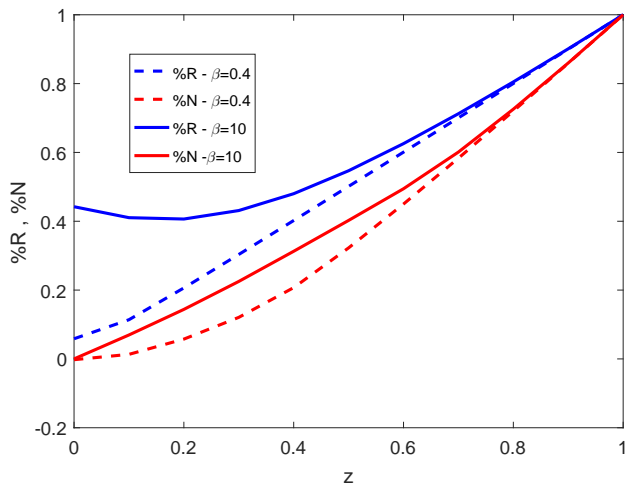}
\caption{Relative Roughness ($\%R$) (blue) and Relative Negativity
($\%N$) (red) as function of $z$ for the state $\rho_{z}$ for
$M=10$, $\beta = 0.4$ (dashed lines) and $\beta = 10$ (full lines).}
\label{fig:RxN_norm}
\end{figure}

\begin{figure}[htb]
\includegraphics[width=\linewidth]{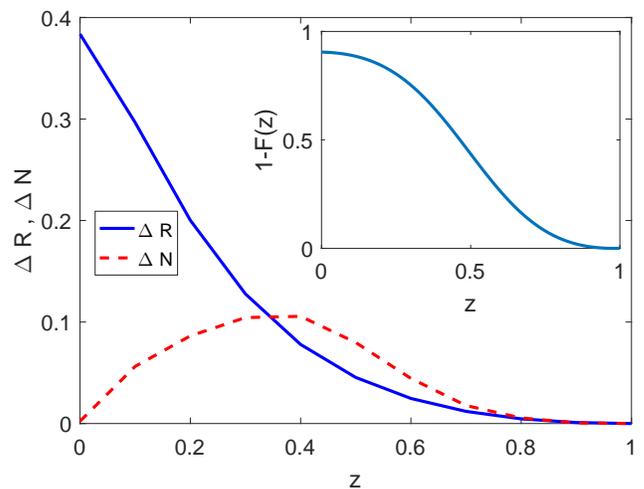}
\caption{Differences $R(\beta=0.4)-R(\beta=10)$ (blue line) and
$N(\beta=0.4)-N(\beta=10)$ (red dashed line) as function of $z$ for
the state $\rho_{z}$ for $M=10$.} \label{fig:RxN_dif}
\end{figure}

Since the Negativity is not bounded to unity, then in order to compare the curves, we show, in
figure \ref{fig:RxN_norm}, the relative Roughness and relative Negativity, respectively defined as
$\%R=R(z)/R(1)$ and $\%N=N(z)/N(1)$. Both measures must find their largest values on $z=1$,
since a Fock state for $M=10$ is more quantum than a thermal state, but $\%R$ is not monotonic on
$z$, specially for large values of $\beta$, as we have already discussed.
We plot these quantities for a large ($\beta=0.4$) and a small temperature ($\beta=10$).
We also studied the differences $\Delta R = R(\beta=0.4)-R(\beta=10)$ and
$\Delta N = N(\beta=0.4)-N(\beta=10)$ in figure \ref{fig:RxN_dif}, which are, respectively, the
differences between $R$ and $N$ at large and small temperatures. For $z=0$ we have a thermal state,
so they must be quite distinct at different temperatures. On the other hand, when $z=1$, $\rho_{z}$
goes to the $M$ Fock state, no matter if the temperature is small or large, so both $\Delta R$ and
$\Delta N$ must be zero.
As we plot both quantities, we see that $\Delta R$ is a monotonic function of $z$, while $\Delta N$
is not. This means that Negativity has failed to discriminate this mixed quantum state.
Again, we appeal to a known quantity, the fidelity $F$ between two quantum states \cite{Vanicek2006},
to support our claim. The fidelity between two mixed states $\rho_{1}$ and $\rho_{2}$ is
defined as  $F=Tr(\rho_{1} \rho_{2})/Tr(\rho_{1}^2)$.
The fidelity between the the states $\rho_{z}(\beta=0.4)$ and $\rho_{z}(\beta=10)$ is monotonic on
$z$, so it is the distance $(1-F)$ between them, as we can see on the inset of figure
\ref{fig:RxN_dif}.
Its behavior is similar to the one that what we observe on $R$. The Negativity, however, is not
monotonic on $z$.

\begin{figure}[h]
\center
\subfigure[ref1][Roughness]{%
\includegraphics[width=\linewidth]{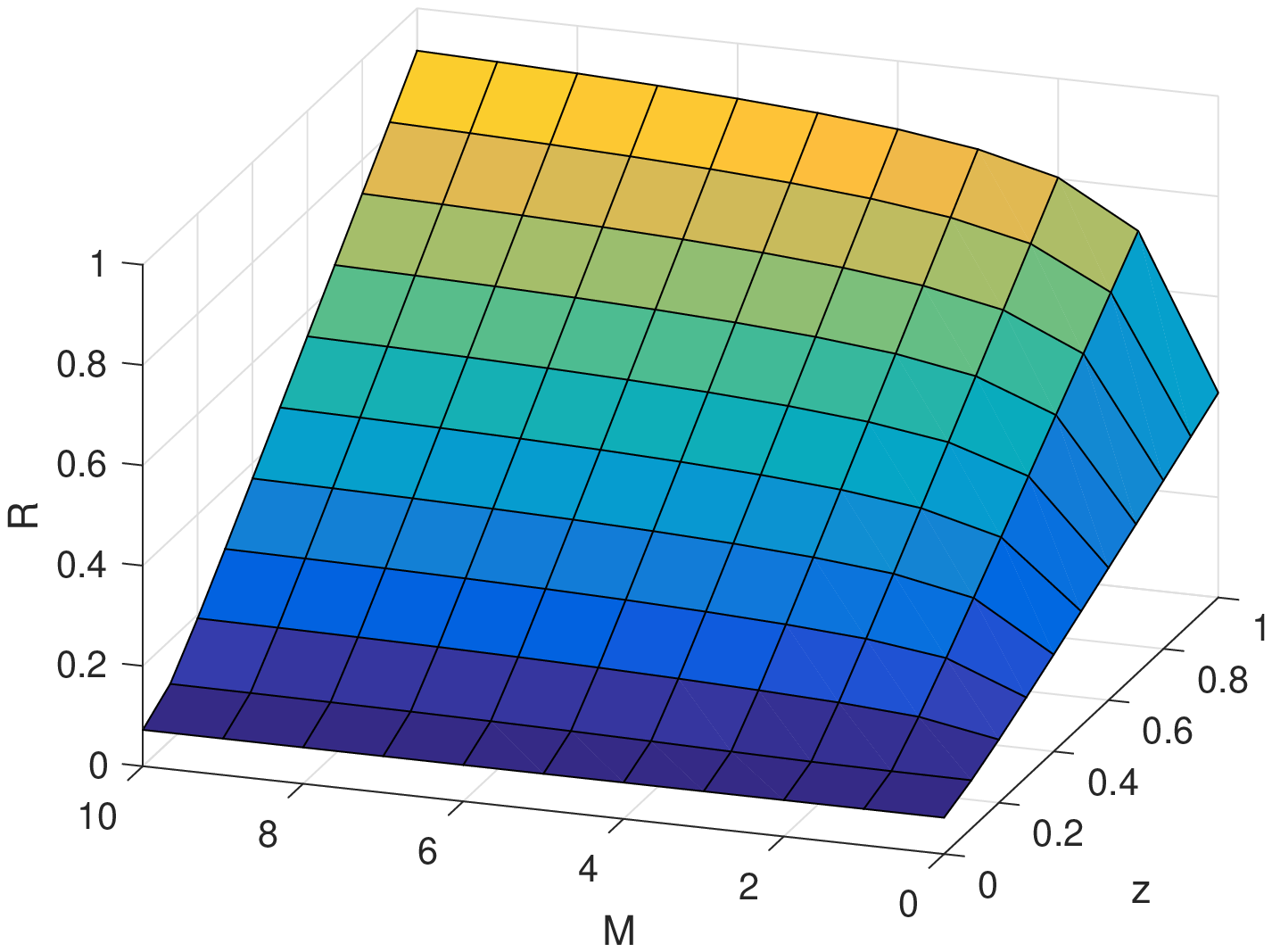}} \qquad %
\subfigure[ref2][Negativity]{%
\includegraphics[width=\linewidth]{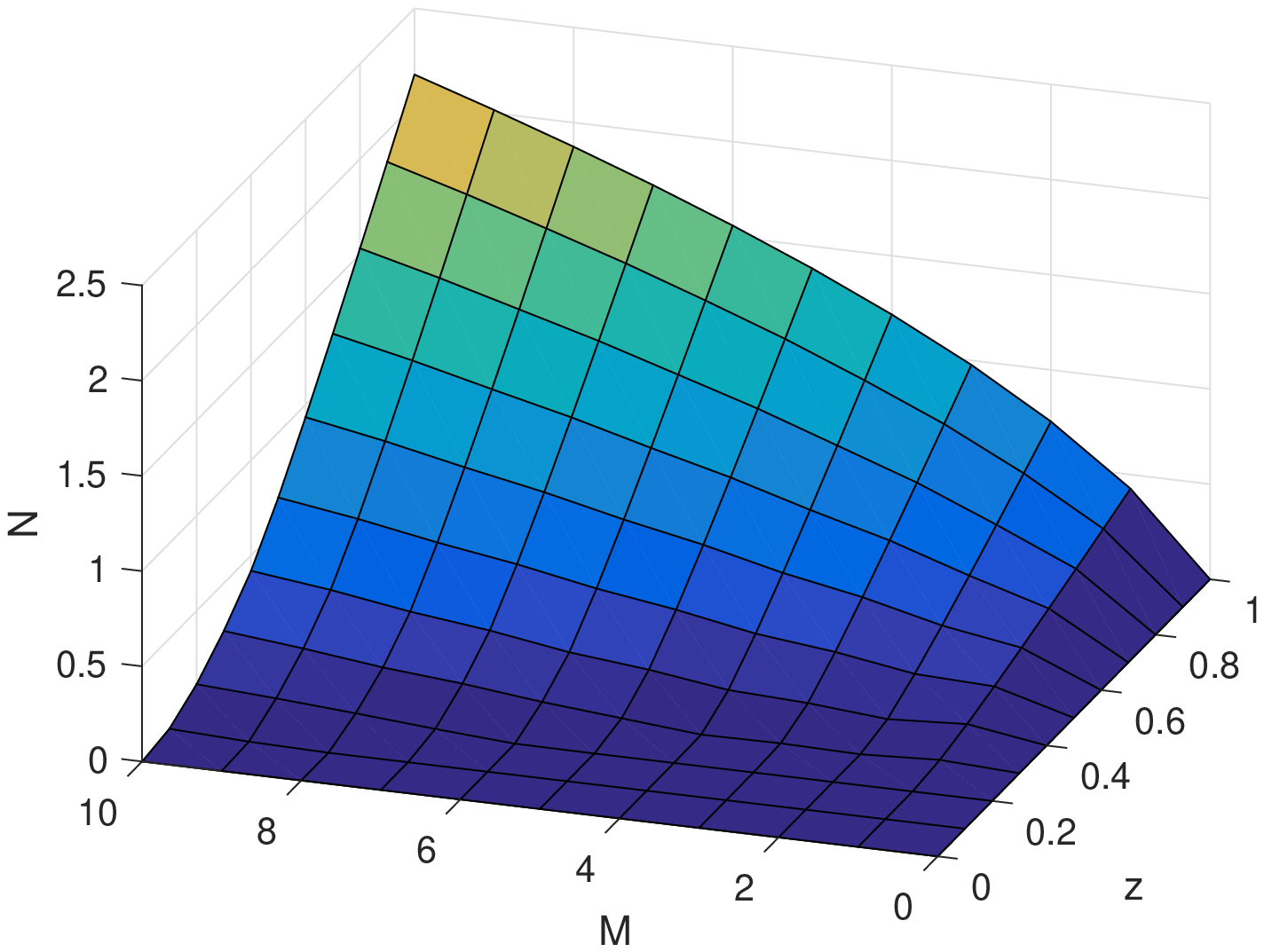}}
\caption{Roughness (a) and Negativity (b) as function of $z$ and $M$
for the state $\rho_{z}$ for $\beta = 0.5$ } \label{TermicoZ3dBetaF}
\end{figure}

\begin{figure}[h]
\center
\subfigure[ref1][Roughness]{%
\includegraphics[width=\linewidth]{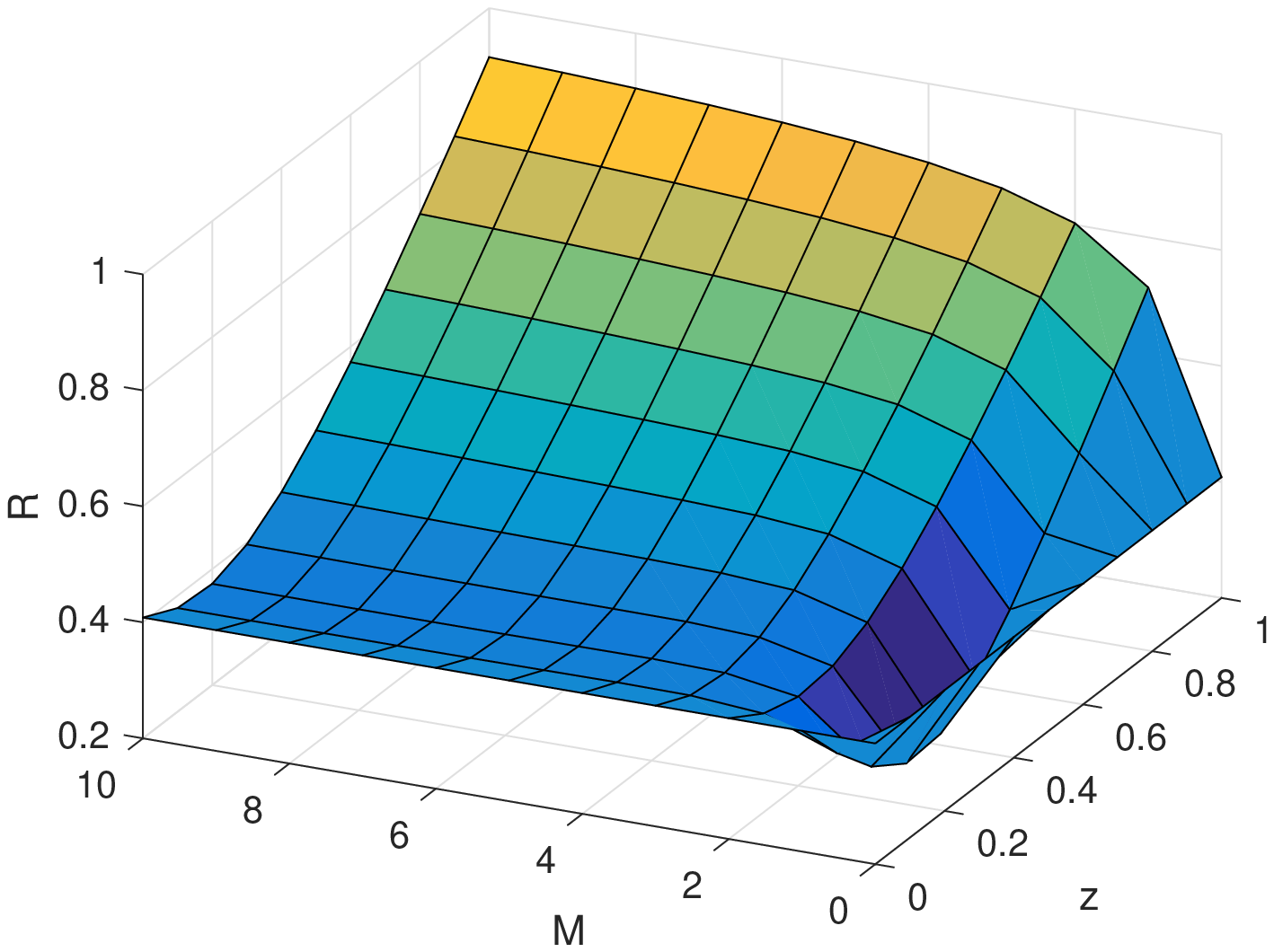}} \qquad %
\subfigure[ref2][Negativity]{%
\includegraphics[width=\linewidth]{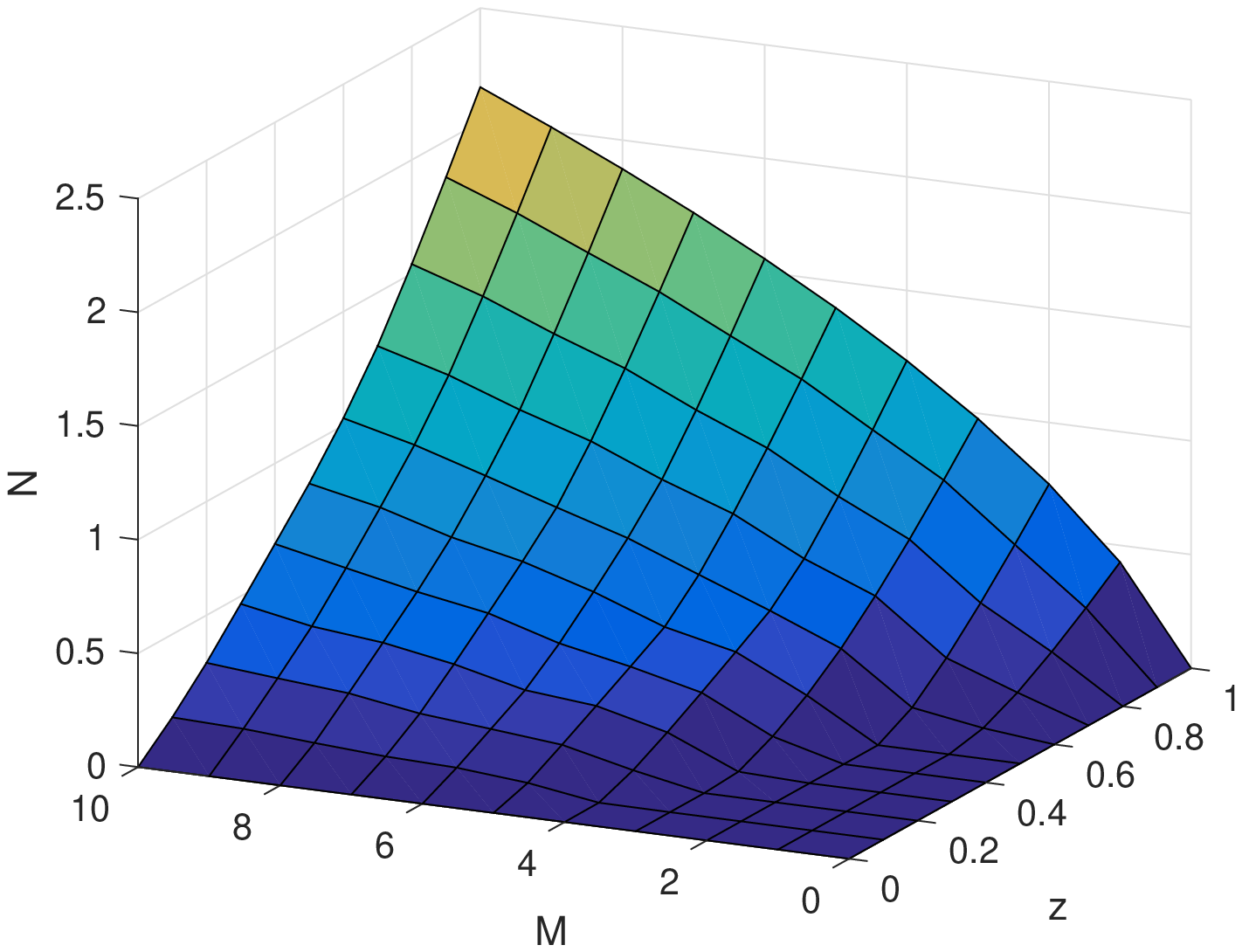}}
\caption{Roughness (a) and Negativity (b) as function of $z$ and $M$
for the state $\rho_{z}$ for $\beta = 10$} \label{TermicoZ3dNFix2}
\end{figure}

\begin{figure}[h]
\center
\subfigure[ref1][Roughness]{%
\includegraphics[width=\linewidth]{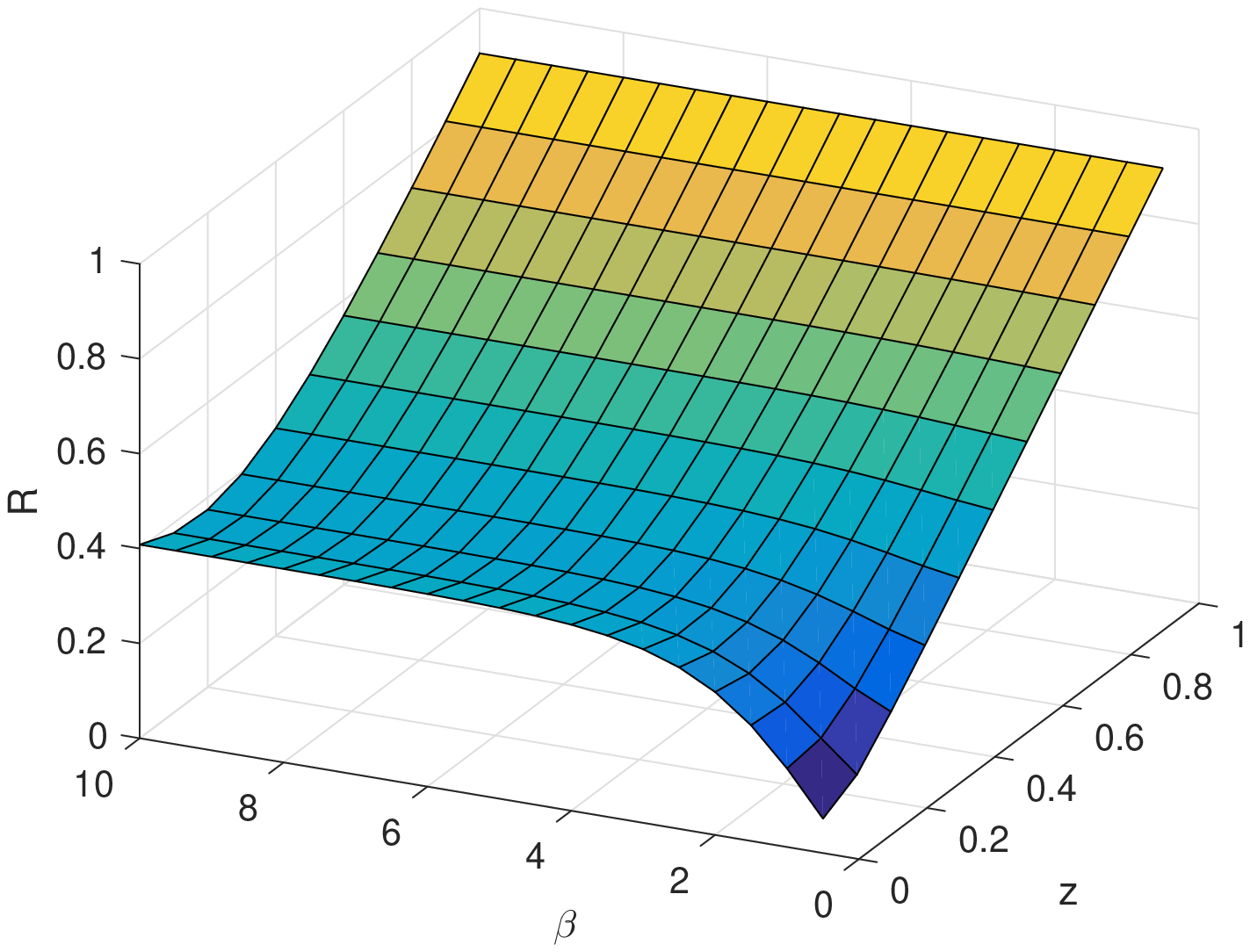}} \qquad %
\subfigure[ref2][Negativity]{%
\includegraphics[width=\linewidth]{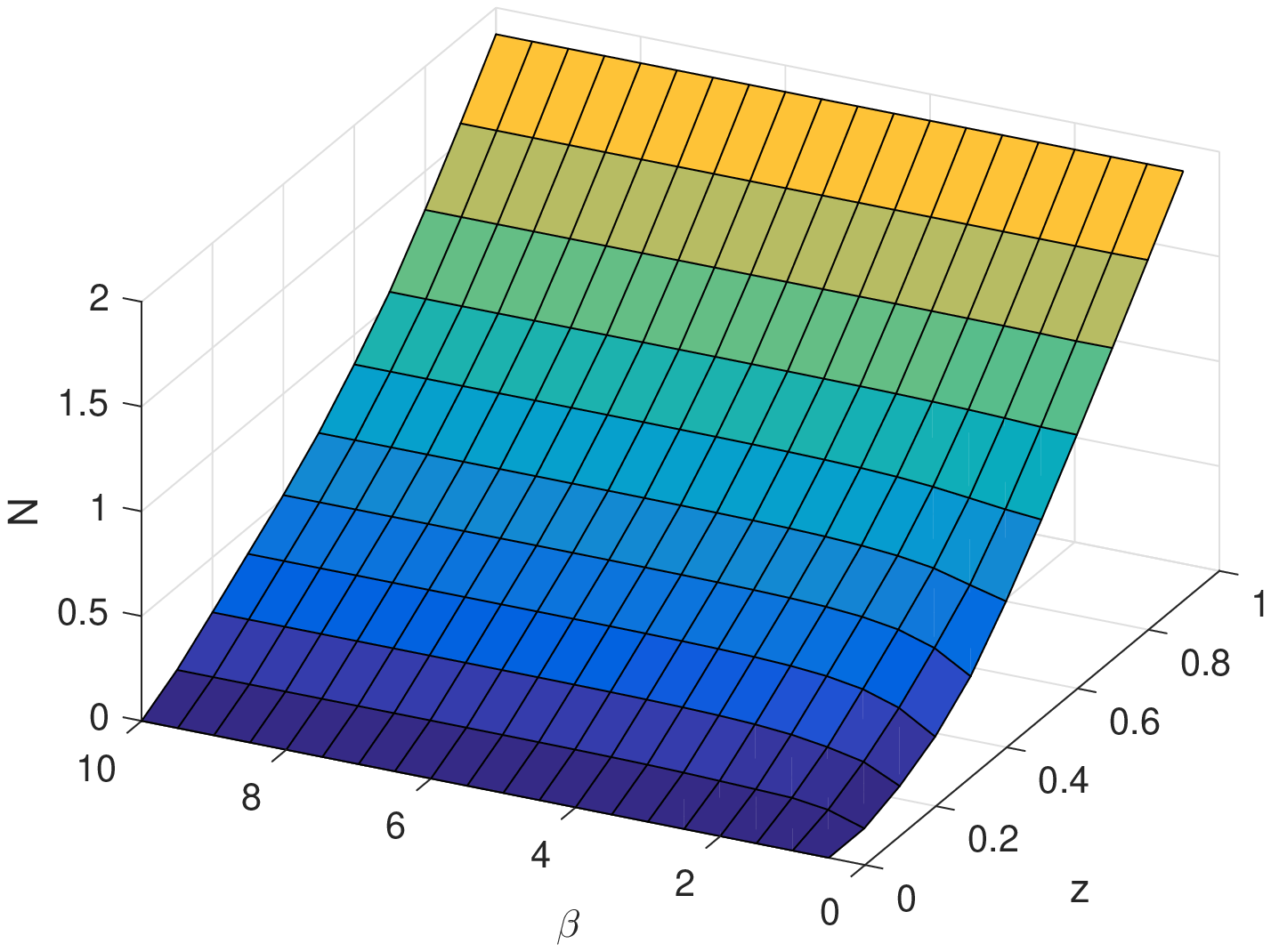}}
\caption{Roughness (a) and Negativity (b) as function of $z$ and
$\beta$ for the state $\rho_{z}$ for $M = 8$.}
\label{TermicoZ3dNFix}
\end{figure}

In Figures \ref{TermicoZ3dBetaF}, \ref{TermicoZ3dNFix2}  and \ref{TermicoZ3dNFix}, we plot graphics
for both $R$ and $N$ as functions of $M$ and $z$. Since that Negativity for thermal states is zero
for any temperature, so $N$ is zero for a large set of states, which is shown in the dark blue part
of figures. The Roughness, on the other hand, is zero only for large temperatures (small $\beta$)
and for $z\approx0$, this is evident on figure \ref{TermicoZ3dNFix}. Comparing figures
\ref{TermicoZ3dBetaF} and \ref{TermicoZ3dNFix2}, we can see that Negativity does not discriminate
mixed states with different temperatures, while the Roughness is sensitive to it.
This fact is most evident in figure \ref{TermicoZ3dNFix}.

\section{Dynamic Distance Measure: the Quartic Model} \label{sec:DDM}

As we have shown, the Roughness is a good measure of how quantum is a state, but it does not tell us
anything about the dynamics, then we also defined the Dynamic Distance Measure $(D)$, which is given
by
\begin{eqnarray} \label{eq:DDM} \label{Distclass}
 D(\Psi (t))&=&\left[ \pi \int \int_{B}\left\vert f(x,p,t)-Q(\Psi
(t))\right\vert ^{2}dxdp\right] ^{1/2}, 
\\
f(x,p,0)&=&Q(x,p,0),  \nonumber
\end{eqnarray}%
where $f$ is the classical Liouville evolution for the corresponding
classical Hamiltonian. The function $D$ was constructed to measure
quantum aspects of dynamics, and then, identical initial states for the
classical and quantum systems must be considered.
 We observe that $D$ is not limited in general, but
if we exclude states that do not respect Heisenberg's uncertain
principle, then $D\in \lbrack 0,1].$ With classical dissipation, $f$
can become delta function and in this case $D\rightarrow \infty $.
Otherwise, $D=1$ only if $\int
\int_{B}f(x,p,t)Q(\Psi (t))dxdp$ $\rightarrow 0,$ this means that $f$ and $%
Q$ are localized in different regions on phase space. A similar
classicality measure was used by Toscano and collaborators
\cite{Toscano2005}. The main difference is that they used the Wigner
function instead of Husimi function on (\ref{eq:DDM}).
Aiming to quantify how much quantum is the dynamics, we believe that it would be better to
use the Husimi function as in our definition, since the Wigner function inherently carries
information of quantumness of the state:
in other words, maybe, in their definition, they are measuring quantumness both for the
state and dynamics at the same time.

In order to investigate the dynamical aspects, we use the quartic
oscillator model (Kerr oscillator), which was the object of many
investigations \cite{adelcio2012,Oliveira09b,Oliveira2006,
Imamoglu1997,Leonski1996, Faria, Leonski2009,Sivakumar2004,
Baghshahi2014,Leonski2014,Kowalewska2008,Oliveira2006,Oliveira2003,Oliveira2014}
with expressive experimental results \cite{Kirchmair2013}. The
Hamiltonian is given by
\begin{equation}
\hat{H}_{0}=\omega \hbar \hat{a}^{\dagger }\hat{a}+\lambda \hbar
^{2}\left( \hat{a}^{\dagger }\right) ^{2}\hat{a}^{2},
\label{quartic}
\end{equation}%
where $\hat{a}$ and $\hat{a}^{\dagger}$ are creation and
annihilation operators, $\omega$ and $\lambda$ are system
parameters.
Given a general initial state $\rho (0)=\sum_{n,m=0}^{\infty }A_{n}A_{m}^{\ast }\left\vert
n\right\rangle \left\langle m\right\vert $, its time evolution is
\begin{equation*}
\rho (t)=\sum_{n,m=0}^{\infty }e^{it(m-n)\left( \omega +\lambda
\hbar \left[ n+m\right] \right) }A_{n}A_{m}^{\ast }\left\vert
n\right\rangle \left\langle m\right\vert
\end{equation*}%
so the Husimi function is
\begin{equation}
Q(\beta )=e^{-\left\vert \beta \right\vert
^{2}}\frac{\sum_{n,m=0}^{\infty }e^{it(m-n)\left( \omega +\lambda
\hbar \left[ n+m\right] \right) }A_{n}A_{m}^{\ast }}{2\pi } \, ,
\end{equation}%
where $\beta =(x+ip)/\sqrt{2}$.
If the initial state is a coherent state $\left\vert \alpha \right\rangle $, then
its Husimi function is
\begin{equation*}
Q(\beta)=g(\alpha,\beta)\left\vert \sum_{n=0}^{\infty }\frac{(\beta
^{\ast }\alpha e^{-it\omega })^{n}}{n!}e^{-it \left(  \omega
n+\lambda \hbar n^{2}\right)}\right\vert^{2},
\end{equation*}%
where
\begin{eqnarray*}
g(\alpha,\beta)&=&\frac{e^{-\left\vert \beta \right\vert
^{2}-\left\vert \alpha \right\vert ^{2}}}{2\pi }.
\end{eqnarray*}%

 The Wigner function is given by

\begin{eqnarray}
W &=&W(\rho ) \nonumber
\\
&=&\sum_{n,m=0}^{\infty }e^{it(m-n)\left( \omega +\lambda \hbar \left[ n+m%
\right] \right) }A_{n}A_{m}^{\ast }W(\left\vert n\right\rangle
\left\langle m\right\vert )
\end{eqnarray}%
where $W(\left\vert n\right\rangle \left\langle m\right\vert ) =
\Pi_{m,n}$ and they are defined in \ref{eq:Pimn}.

\subsection{Classical Liouville evolution}

The classical equivalent Hamiltonian is
\cite{Oliveira2003,Oliveira2014}

\begin{equation}
H_{cl}=\omega \hbar \left\vert \alpha \right\vert ^{2}+\lambda \hbar
^{2}\left\vert \alpha \right\vert ^{4}.
\end{equation}%
At the initial time we have \qquad \qquad \qquad\
\begin{equation}
f_{t}(\alpha ,\beta (x,y),0)=\frac{1}{2\pi }\exp \left[ -\left\vert
\beta -\alpha \right\vert ^{2}\right] ,
\end{equation}%
then we have \cite{Oliveira2006}

\begin{equation*}
f_{t}(x,p,t)=\frac{1}{2\pi }\exp \left[ -\left\vert \alpha -\frac{x+ip%
}{\sqrt{2}}\exp \left[ it\left( \omega +\lambda \left(
x^{2}+p^{2}\right) \right) \right] \right\vert ^{2}\right] .
\end{equation*}

\begin{figure}[htb]
\includegraphics[width=\linewidth]{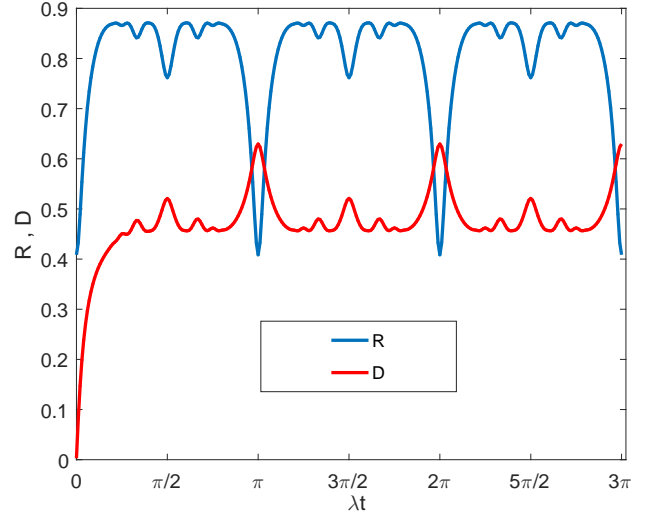}
\caption{Dynamic Distance Measure ($D$, red line) and Roughness ($R$, blue line) as
function of time for the Quartic Oscillator with a coherent initial
state $\alpha =2$ and $\omega=0$.} \label{DxT dif}
\end{figure}

\begin{figure}[htb]
\includegraphics[width=\linewidth]{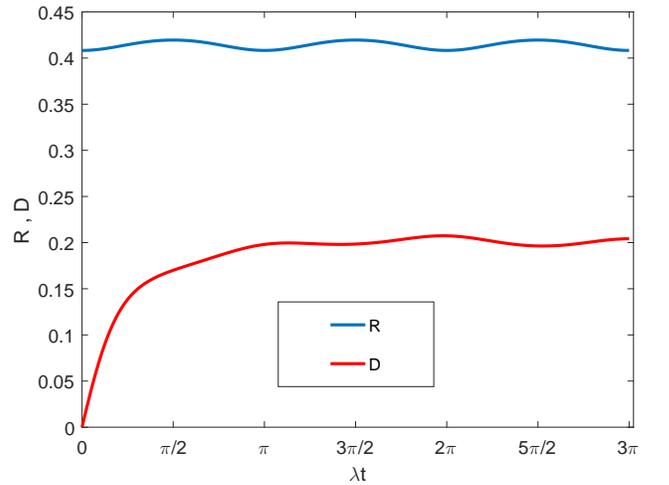}
\caption{Dynamic Distance Measure ($D$, red line) and Roughness ($R$, blue line) as
function of time for the Quartic Oscillator with a coherent initial
state $\alpha= 0.3$ and $\omega=0$.} \label{DxT_dif2}
\end{figure}

In figures \ref{DxT dif} and \ref{DxT_dif2} we show the Roughness
and the Dynamic Distance Measure (DDM) as function of time, respectively
for $\alpha =2$ and $\alpha =0.3$. As we can see, the quantum
aspects of the dynamics are amplified as we increase the classical
action ($S$), since  $S \propto \mid\alpha\mid^{2}$, as was
previously observed \cite{Oliveira2003}. This feature is attenuated
as an environment is included \cite{Oliveira2006} and also when the
system is monitored \cite{Oliveira2014}.

From our numerical simulations, we conjecture, for the Kerr oscillator, that Roughness and DDM have a
complementary aspect in the permanent regime, as Roughness increases DDM decreases and vice-versa.
On the other hand, we could not find a simple mathematical relationship between them that would hold
for any model.

As a final remark, we emphasize that a non-zero Roughness does not mean that the state is quantum,
but it has some quantum characteristic, which is more detectable as Roughness gets closer to its
upper bound. For example, we can have a quantum system in a thermal state with high temperature,
expected to behave as classical, but with discrete spectrum.
On the other hand, a maximum Roughness does not mean that all observable will necessarily have
experimental results that diverge from its classic counterpart, but that such probability of
detection is maximum. Indeed, that is the case of the Kerr oscillator:
for the position expectation value it behaves almost classically, until the revival time
$t_{r}\approx \pi/(2 \lambda)$ \cite{Oliveira2006}, but as we can see in figure \ref{DxT dif}, the
Roughness has already reached its maximum for times of the order of $\pi/(4 \lambda)$.
Moreover, the fact that a system is classical or not depends on the choice of the observable
\cite{Oliveira2003} and how the measurement is performed
\cite{Jacobs,Kofler,renato2007,Ball1998,AngeloLow,Milburn,adelcio2012,Oliveira2013,Oliveira2014}.

\section{Conclusion}
Inspired by the usual definition of roughness, we define the Roughness measure of a state as a
distance measure between its Wigner and Husimi functions. As a general result, we emphasize that the
Roughness has proved to be an effective measure for characterization of states, being able to
discriminate pure states and mixed states. Because the Roughness is bounded, it is possible to compare
distinct states by quantifying the degree of their classicality. The Roughness of a state lies in
the interval [0,1], so we can say that the state is more classical as its roughness is closer to
zero, while it is more quantum as it approaches the unity. The degree of classicality of a state is,
in this way, a fuzzy-like measure, and only limit states can be said fully classic or quantum, the
others have a degree of classically that varies continuously between the extremes. Among the states
approaching the upper bound, we show that the pure Fock state $\left| n \right\rangle$ at the limit
$\lim n\to \infty$ has maximum  Roughness, as well as the squeezed state at the limits of maximum
compression. The minimum Roughness value is reached for mixed states at the infinite entropy
boundary. Comparing Roughness with Negativity, we believe that Roughness does a better job
distinguishing between two given states, since Negativity is zero for any positive state, while
Roughness is non-zero for most of them. With the results on \ref{app:FockGen}, we can analytically
evaluate the Roughness for any state that can be represented using the Fock states as a basis.
We also investigated the dynamics of the Roughness for the quartic oscillator model, and we observed
that, for the quartic oscillator, there is a certain complementarity relationship between Roughness
and the Distance between quantum and classical Liouvillian dynamics.

\section*{Acknowledgments}
ACO, HCLF and ACLA gratefully acknowledge the support of Brazilian agency Funda\c{c}\~ao de Amparo
\`a Pesquisa do Estado de Minas Gerais (FAPEMIG) through grant No. APQ-01366-16.
BA acknowledges financial support from the Brazilian ministries MEC and MCTIC, 
CNPq (Grant No. 444927/2014-0) and INCT.

 \appendix
\section{An auxiliary result}

While we were proving the upper bound (\ref{eq:Rbounds}) for the Roughness, we found an auxiliary
result. 
For any given quantum state, we rewrite Roughness as
\begin{equation} \label{eq:ap1}
R^{2} = R_{W^{2}}^{2}+R_{Q^{2}}^{2}-R_{WQ}^{2} \, ,
\end{equation}
in the same sense as we did on equations (\ref{eq:R2W2def}), (\ref{eq:R2Q2def}) and (\ref{eq:R2WQdef})
for a Fock state.
First and second terms on the RHS of the equality are clear positive, but we were not sure about the
last one, since the Wigner function is not necessarily positive.
Using once again the Fourier transform, as we have used to prove the upper bound, we get
\begin{eqnarray}
R_{WQ}^{2} &:=&
\int_{\R^{2}}dq\,dp\,W(q,p)Q(q,p) = (W,Q)= (\widehat{W},\widehat{Q}) =
\nonumber  \\
&=& \int_{\R^{2}}du\,dv\,e^{-\frac{1}{4}(u^{2}+v^{2})}\left\vert \hat{W%
}(u,v)\right\vert ^{2}\geq 0, \label{eq:RWHgeneral}
\end{eqnarray}%
which proves that every single term on (\ref{eq:ap1}) is positive.
As consequence, we have an upper bound for Roughness
\begin{equation*}
R^{2}\leq 2\pi \int_{\Re ^{2}}dq\,dp\,\left( W(q,p)^{2}+Q(q,p)^{2}\right) ,
\end{equation*}
which may be useful sometimes.

\section{Roughness for a Fock state, analytical results} 
\label{app:Fock}

We give in this appendix some details for the results obtained on subsection \ref{ssec:Fock}.
We start proving the first limit on (\ref{eq:limits}). For this, we use Stirling's approximation
\cite{Shiryaev1995}
\begin{equation} \label{eq:Stir}
\sqrt{2\pi n} \left( \frac{n}{e} \right)^n \leq n! \leq e^{\frac{1}{12n}} \sqrt{2\pi n} \left( \frac{n}{e} \right)^n .
\end{equation}
Using it properly on (\ref{eq:R2H2}), we get
\begin{equation} \label{eq:CotaS2n}
\frac{e^{-\frac{1}{6n}}}{\sqrt{\pi n}}   \leq  \frac{1}{2^{2n}} {2n \choose n} \leq
\frac{e^{\frac{1}{24n}}}{\sqrt{\pi n}}   ,
\end{equation}
and since left and right sides of the inequality above goes to 0 as $n \to \infty$, then we finish
demonstration by the squeeze theorem.

Now we look to (\ref{eq:R2WH_F}) and rewrite it as
\begin{equation} \label{eq:R2WH_Cn}
R^2_{W_nQ_{n}} = \frac{4}{3} \left(\frac{1}{3}\right)^n \underbrace{(-1)^n \sum_{j=0}^n
\frac{(n+j)!}{(j!)^2(n-j)!} \left( -\frac{4}{3} \right)^j}_{=:C_n}\, ,
\end{equation}
where we just have defined $C_n$.
It is not clear above that $C_n > 0$, as it must be\footnote{Actually, since (\ref{eq:RWHgeneral})
was obtained for a general Wigner function, it guarantees that $C_n>0$. Anyway it is worthwhile to
explicitly show this in order to prove (\ref{eq:ineq}), since it will be useful to evaluate the
second limit on (\ref{eq:limits}).}
from (\ref{eq:RWHgeneral}).

We rewrite $C_n$ as
\begin{eqnarray*}
C_n &=& \frac{(-1)^n}{n!} \sum_{j=0}^n (j+n)\ldots(j+1) 
\frac{j! \, n!}{(j!)^{2}(n-j)!} \left( -\frac{4}{3} \right)^j =
\\
&=& \frac{(-1)^n}{n!} \sum_{j=0}^n \left( \sum_{k=0}^n {n+1 \brack k+1} j^k \right)
{n \choose j} \left( -\frac{4}{3} \right)^j,
\end{eqnarray*}
where ${n \brack k}$ are the unsigned Stirling numbers of the first kind \cite{Graham1994}, and
they appear as coefficients for the polynomials on $j$ from the product above. Indeed, Stirling
numbers of the first kind appear on rising factorials definition
$$
x^{(n)} := x(x+1)\ldots(x+n-1)  \Rightarrow
x^{(n)} = \sum_{k=0}^n {n \brack k} x^k\, ,
$$
and many other applications.
We reverse the order of summation to get
$$
C_n = \frac{(-1)^n}{n!} \sum_{k=0}^n {n+1 \brack k+1} \left[ \sum_{j=0}^n j^k
{n \choose j} \left( -\frac{4}{3} \right)^j \right]\, .
$$
The summation on index $j$ above may be written as derivatives of binomial as
\begin{equation} \label{eq:CnAux}
C_n = \frac{(-1)^n}{n!} \sum_{k=0}^n {n+1 \brack k+1} \left.
\left( x \frac{d}{dx} \right)^k (1+x)^n \right|_{x=-\frac{4}{3}},
\end{equation}
and we may show that
$$
\left( x \frac{d}{dx} \right)^k (1+x)^n = \sum_{j=0}^k {k \brace j} \frac{n!}{(n-j)!}
x^j (1+x)^{n-j}\, ,
$$
where ${n \brace j}$ are the Stirling numbers of second kind \cite{Graham1994}.
We replace expression above on (\ref{eq:CnAux}), and reversing summation we get
\begin{widetext}
\begin{eqnarray}
C_n &=& 
(-1)^n \sum_{j=0}^n \left( \sum_{k=j}^n {n+1 \brack k+1} {k \brace j} \right)
\frac{1}{(n-j)!} \left( -\frac{4}{3} \right)^j \left( -\frac{1}{3} \right)^{n-j}  =
\nonumber \\
\label{eq:Cn}
&=& \left( \frac{1}{3} \right)^{n} \sum_{j=0}^n \left( \sum_{k=j}^n {n+1 \brack k+1}
{k \brace j} \right) \frac{1}{(n-j)!} \,  4^j  >0,
\end{eqnarray}
\end{widetext}
and so we proved that $C_n>0$. But, even better, we were able to prove that
\begin{equation} \label{eq:S1S2}
\sum_{k=j}^n {n+1 \brack k+1} {k \brace j} = (n-j)! {n \choose j}^2,
\end{equation}
and so we can find (\ref{eq:R2WH_2}).

To continue studying $R^2_{W_nQ_{n}}$ properties, we match eq. (\ref{eq:R2WH_2}) to a polynomial
$\mathcal{P}_n(t)$ whose coefficients are the square of binomial coefficients \cite{Fetecau2011},
namely
\begin{eqnarray} 
\mathcal{P}_n(t) &:=&
\frac{1}{2\pi} \int_{0}^{2\pi}d\theta \,  \big( 1 + t^2 - 2t \cos \theta \big)^n =
\nonumber \\
\label{eq:Pn}
&=& \sum_{j=0}^n {n \choose j}^2 t^{2j} \, ,
\end{eqnarray}
so we get
\begin{eqnarray*}
R^2_{W_nQ_{n}} &=& \frac{4}{3} \left( \frac{1}{9} \right)^n \mathcal{P}_n(2) =
\\
&=&
\frac{4}{3} \left( \frac{1}{9} \right)^n \frac{1}{2\pi} \int_{0}^{2\pi}d\theta \,  \big( 5 - 4 \cos \theta \big)^n =
\\
&=& \frac{4}{3} \left( \frac{1}{9} \right)^n \frac{1}{2\pi} \int_{0}^{2\pi}d\theta \, \left(
 9 \sin^{2}  \frac{\theta}{2}   +  \cos^{2}  \frac{\theta}{2} \right)^n >
\\
&>& \frac{4}{3} \left( \frac{1}{9} \right)^n \frac{9^n}{\pi} \int_{0}^{\pi}d\theta \,
 \sin^{2n} \theta   \, ,
\end{eqnarray*}
Using the known fact that
$$
\frac{1}{\pi} \int_{0}^{\pi} d\theta \, \sin^{2n} \theta = \frac{1}{2^{2n}} {2n \choose n} \, ,
$$
we have
\begin{equation} \label{eq:order}
R^2_{W_nQ_{n}} > \frac{4}{3} \frac{1}{2^{2n}} {2n \choose n} >
\frac{1}{2} \frac{1}{2^{2n}} {2n \choose n} = R^2_{Q_{n}^2} ,
\end{equation}
which proves (\ref{eq:ineq}).
Moreover, we have a lower bound for $R^2_{W_nQ_{n}}$.
It is quite more technical, but we can find a similar upper bound for $R^2_{W_nQ_{n}}$. Indeed, we
find a constant $B>1$, such that
$$
R^2_{W_nQ_{n}} < B \ \frac{4}{3} \frac{1}{2^{2n}} {2n \choose n} ,
$$
and so, again by using squeeze theorem, we prove that $R^2_{W_nQ_{n}} \to 0$.
To do so, we define for each $n$ the constant $B_n$ as
\begin{eqnarray*}
B_n &:=& \left[ \frac{1}{2^{2n}} {2n \choose n} \right]^{-1} 9^{-n} \mathcal{P}_n(2) =
\\
&=& 2^{2n} {2n \choose n}^{-1} \frac{1}{\pi}
\int_0^{\pi} d\theta \, \left( \sin^2 \theta + \frac{1}{9} \cos^2 \theta \right)^n  .
\end{eqnarray*}
Explicit evaluation shows that $B_0=1$ and $B_1=10/9$. 
After some tedious calculations, we can show that, for any $n \geq 1$, we have
$1 \leq B_{n+1} < B_{n}$, and so, we can conclude that $B_n \leq 10/9$, for any $n$, which ends
our proof.

\section{Integrals of $\Pi_{n,m}$ and $\Psi_{n,m}$}
\label{app:FockGen}

We evaluate here useful quantities to find the Roughness for general states like that on Eq.
(\ref{eq:FockGen}). Since the Wigner transform is linear \cite{Wigner1932}, the Wigner function for
this state is
\begin{equation} \label{eq:Wmn}
W(q,p) = \sum_{m,n=0}^{\infty} A_{n,m} \Pi_{m,n} (\alpha)\ ,
\end{equation}
where $\alpha = (q+ip)/\sqrt{2}$, and $\Pi_{m,n} (\alpha)$ is given by
\begin{widetext}
\begin{equation} \label{eq:Pimn}
\Pi_{m,n} (\alpha)= \begin{cases}
\displaystyle{\frac{(-1)^m}{\pi} \sqrt{\frac{m!}{n!}} e^{-2|\alpha|^2} (2\alpha)^{n-m} L_{m}^{n-m}(4|\alpha|^2)} \, ,
 & \mbox{if } n\geq m,
\\
\\
\displaystyle{\frac{(-1)^n}{\pi} \sqrt{\frac{n!}{m!}} e^{-2|\alpha|^2} (2\alpha)^{m-n} L_{n}^{m-n}(4|\alpha|^2)} \, ,
 & \mbox{if } n < m.
\end{cases}
\end{equation}
\end{widetext}
The $L_{m}^{n-m}$ are the associated Laguerre functions
\cite{Arfken2011}. Analogously, Husimi function for
(\ref{eq:FockGen}) is
\begin{equation} \label{eq:Qmn}
Q(q,p) = \sum_{m,n=0}^{\infty} A_{n,m} \Psi_{m,n} (\alpha)\ ,
\end{equation}
where
\begin{equation} \label{eq:Psimn}
\Psi_{m,n} (\alpha) = \frac{\alpha^n \, (\alpha^{\ast})^m}{2\pi \sqrt{n!m!}} e^{-|\alpha|^2},
\end{equation}
where $\alpha^{\ast}$ denotes the complex conjugate.
It is important to stress that, for $n=m$, equations (\ref{eq:Pimn}) and (\ref{eq:Psimn}),
respectively give us functions (\ref{eq:WFock}) and (\ref{eq:HFock}) for pure states.

From definitions (\ref{eq:Wmn})-(\ref{eq:Psimn}), the Roughness for general state (\ref{eq:FockGen})
is
\begin{widetext}
\begin{eqnarray} \label{eq:RGen}
R^2 &=& 2\pi \int_{\mathbb{R}^2}dq\, dp\, \big| W(q,p) - Q(q,p) \big|^2 =
\\
&=&
\sum_{n,m,n',m'} A_{n,m}^{\ast} A_{n',m'} \left[
R_{\Pi_{m,n} \Pi_{m',n'}}^{2} + R_{\Psi_{m,n} \Psi_{m',n'}}^{2} -
\left( R_{\Pi_{m,n} \Psi_{m',n'}}^{2} + R_{\Psi_{m,n} \Pi_{m',n'}}^{2} \right) \right],
\nonumber
\end{eqnarray}
\end{widetext}
where we define
\begin{subequations}
\begin{eqnarray} \label{eq:a}
R_{\Pi_{m,n} \Pi_{m',n'}}^{2} &=& 2\pi \int_{\mathbb{R}^2}dq\, dp\, \Pi_{m,n}^{\ast} \Pi_{m',n'}\, ,
\\
R_{\Psi_{m,n} \Psi_{m',n'}}^{2} &=& 2\pi \int_{\mathbb{R}^2}dq\, dp\, \Psi_{m,n}^{\ast} \Psi_{m',n'}\, ,
\\
R_{\Pi_{m,n} \Psi_{m',n'}}^{2} &=& 2\pi \int_{\mathbb{R}^2}dq\, dp\, \Pi_{m,n}^{\ast} \Psi_{m',n'}\, ,
\\
\label{eq:d}
R_{\Psi_{m,n} \Pi_{m',n'}}^{2} &=& 2\pi \int_{\mathbb{R}^2}dq\, dp\, \Psi_{m,n}^{\ast} \Pi_{m',n'}\, .
\end{eqnarray}
\end{subequations}

We now present results for the integrals (\ref{eq:a})-(\ref{eq:d}) above, which are necessary to
compute Roughness for general states (\ref{eq:FockGen}). All these computations were performed
analytically, using some well known properties of associated Laguerre functions
\cite{Gradshteyn2007}.

We obtain
\begin{widetext}
\begin{eqnarray} \label{eq:PiPi}
R_{\Pi_{m,n} \Pi_{m',n'}}^{2} &=& \delta_{n,n'} \delta_{m,m'} \, .
\\
R_{\Psi_{m,n} \Psi_{m',n'}}^{2}  &=& \frac{\delta_{n-m,n'-m'}}{\sqrt{n!m!n'!m'!}}
\left(\frac{1}{2}\right)^{\frac{n+m+n'+m'}{2}+1} \left( \frac{n+m+n'+m'}{2} \right)! \, .
\end{eqnarray}
\end{widetext}
Albeit the other two integrals are obtained in the same way, their expressions are a little bit more
complicated. First we must define $X:=\max (n,m)$, $Y:=\min (n,m)$, and similar
quantities for prime indices. We get
\begin{widetext}
\begin{eqnarray} \label{eq:PiPsi}
 R_{\Pi_{m,n} \Psi_{m',n'}}^{2} &=& \frac{2}{3} \delta_{n-m,n'-m'}
(-1)^Y \sqrt{\frac{Y!}{X!X'!Y'!}} 2^{X-Y} \left( \frac{1}{3} \right)^{\frac{X-Y+X'+Y'}{2}} 
\sum_{j=0}^Y {X \choose Y-j} \frac{\left( \frac{X-Y+X'+Y'}{2} + j \right)!}{j!}
\left(-\frac{4}{3}\right)^j\ ,
\\
\label{eq:PsiPi}
 R_{\Psi_{m,n} \Pi_{m',n'}}^{2} &=& \frac{2}{3} \delta_{n-m,n'-m'}
(-1)^{Y'} \sqrt{\frac{Y'!}{X!Y!X'!}} 2^{X'-Y'} \left( \frac{1}{3} \right)^{\frac{X+Y+X'-Y'}{2}} 
\sum_{j=0}^{Y'} {X' \choose Y'-j} \frac{\left( \frac{X+Y+X'-Y'}{2} + j \right)!}{j!}
\left(-\frac{4}{3}\right)^j\ .
\end{eqnarray}
\end{widetext}
One must see that (\ref{eq:PiPsi}) and (\ref{eq:PsiPi}) are the same expression with non-prime and
prime indices exchanged.

It may be not easy to evaluate the Roughness by hand for a general state (\ref{eq:FockGen}) using
equations (\ref{eq:PiPi})-(\ref{eq:PsiPi}), but we emphasize that we have analytically calculated
integrals on Eq. (\ref{eq:RGen}), and it will certainly save lots of computational resources on this
task. It is easier for a computer to numerically evaluate sums like those presents on
(\ref{eq:PiPi})-(\ref{eq:PsiPi}) than to compute integrals like (\ref{eq:RGen}).



\end{document}